\documentclass[a4paper,11pt]{article}
\pdfoutput=1
\usepackage{jheppub}
\usepackage{amsmath}
\usepackage{float}
\usepackage{soul}
\usepackage{color}

\newcommand{\ii}{\mathrm{i}}

\newcommand{\Tr}{\mathrm{Tr}\,}

\newcommand{\cW}{{\mathcal{W}}}

\newcommand{\one}{{\rm 1\kern -.9mm l}}

\newcommand{\be}{\begin{equation}}
\newcommand{\ee}{\end{equation}}

\newdimen\tableauside\tableauside=1.0ex
\newdimen\tableaurule\tableaurule=0.4pt
\newdimen\tableaustep
\def\phantomhrule#1{\hbox{\vbox to0pt{\hrule height\tableaurule
width#1\vss}}}
\def\phantomvrule#1{\vbox{\hbox to0pt{\vrule width\tableaurule
height#1\hss}}}
\def\sqr{\vbox{%
  \phantomhrule\tableaustep
\hbox{\phantomvrule\tableaustep\kern\tableaustep\phantomvrule\tableaustep}%
  \hbox{\vbox{\phantomhrule\tableauside}\kern-\tableaurule}}}
\def\squares#1{\hbox{\count0=#1\noindent\loop\sqr
  \advance\count0 by-1 \ifnum\count0>0\repeat}}
\def\tableau#1{\vcenter{\offinterlineskip
  \tableaustep=\tableauside\advance\tableaustep by-\tableaurule
  \kern\normallineskip\hbox
    {\kern\normallineskip\vbox
      {\gettableau#1 0 }%
     \kern\normallineskip\kern\tableaurule}%
  \kern\normallineskip\kern\tableaurule}}
\def\gettableau#1 {\ifnum#1=0\let\next=\null\else
  \squares{#1}\let\next=\gettableau\fi\next}
\def\XXint#1#2#3{{\setbox0=\hbox{$#1{#2#3}{\int}$}
     \vcenter{\hbox{$#2#3$}}\kern-.5\wd0}}

\usepackage[usenames,dvipsnames,svgnames,table]{xcolor}

\usepackage{tikz}
\usetikzlibrary{shapes.symbols,shapes.geometric,shapes.misc,circuits.logic.US}
\usetikzlibrary{matrix}
\usetikzlibrary{graphs}
\usetikzlibrary{trees}
\usetikzlibrary{arrows,arrows.spaced}
\usetikzlibrary{decorations.markings,patterns}
\usetikzlibrary{backgrounds,fit}

\tikzstyle{gauge} = [circle, text centered, draw=black, minimum height=1.2cm]
\tikzstyle{squashedflavor} = [rectangle, text centered, draw=black, minimum height=1cm,minimum width=1.2cm]
\tikzstyle{flavor} = [rectangle, text centered, draw=black, minimum height=1.2cm,minimum width=1.2cm]
\tikzstyle{oldgaugedflavor} = [and gate,draw, point up, minimum height=1cm,draw=black]
\tikzstyle{gaugeS} = [circle, text centered, draw=black, minimum height=6ex]
\tikzstyle{flavorS} = [rectangle, text centered, draw=black,minimum height=6ex,minimum width=6ex]

\makeatletter
\tikzset{arc style/.initial={}}
\pgfdeclareshape{dash circled box}{
    \inheritsavedanchors[from={rectangle}]
    \inheritsavedanchors[from={circle}]
    \inheritanchorborder[from={rectangle}] 
    
    \inheritanchor[from={circle}]{center}
    \inheritanchor[from={rectangle}]{center}
    \inheritanchor[from={rectangle}]{south}
    \inheritanchor[from={rectangle}]{west}
    \inheritanchor[from={rectangle}]{north}
    \inheritanchor[from={rectangle}]{east}

    \backgroundpath{
    \southwest \pgf@xa=\pgf@x \pgf@ya=\pgf@y
    \northeast \pgf@xb=\pgf@x \pgf@yb=\pgf@y
    \pgf@xc=\pgf@xb \advance\pgf@xc by-5pt 
    \pgf@yc=\pgf@yb \advance\pgf@yc by-5pt
    
    \pgf@xx=\pgf@xa \advance\pgf@xx by10pt 
    \pgf@yy=\pgf@ya \advance\pgf@yy by10pt

    \pgfpathmoveto{\pgfpoint{\pgf@xa}{\pgf@ya}}
    \pgfpathlineto{\pgfpoint{\pgf@xa}{\pgf@yb}}
    \pgfpathlineto{\pgfpoint{\pgf@xc}{\pgf@yb}}
    \pgfpathlineto{\pgfpoint{\pgf@xb}{\pgf@yc}}
    \pgfpathlineto{\pgfpoint{\pgf@xb}{\pgf@ya}}
    \pgfpathclose
    \pgfpathmoveto{\pgfpoint{\pgf@xc}{\pgf@yb}}
    \pgfpathlineto{\pgfpoint{\pgf@xc}{\pgf@yc}}
    \pgfpathlineto{\pgfpoint{\pgf@xb}{\pgf@yc}}
    \pgfpathlineto{\pgfpoint{\pgf@xc}{\pgf@yc}}
    
    \pgfpathcircle{\pgfpoint{\pgf@xx}{\pgf@yy}}{\pgf@xb} 
 }
}
\makeatother

\makeatletter
\pgfdeclareshape{barn}{
    \inheritsavedanchors[from={rectangle}]
    \inheritanchorborder[from={rectangle}] 
     \inheritanchor[from={rectangle}]{north}
    \inheritanchor[from={rectangle}]{center}
    \inheritanchor[from={rectangle}]{south}
    \inheritanchor[from={rectangle}]{west}
    \inheritanchor[from={rectangle}]{east}

    \inheritbackgroundpath[from={rectangle}]
   
    \backgroundpath{
    \southwest \pgf@xa=\pgf@x \pgf@ya=\pgf@y
    \northeast \pgf@xb=\pgf@x \pgf@yb=\pgf@y
    \pgf@xc=\pgf@xb \advance\pgf@xc by-\pgf@xa     
    \pgf@yc=\pgf@yb \advance\pgf@yc by-\pgf@ya    
    \advance\pgf@yb by-0.5\pgf@yc
    \pgfpathmoveto{\pgfpoint{\pgf@xa}{\pgf@ya}}
    \pgfpathlineto{\pgfpoint{\pgf@xa}{\pgf@yb}}
    \pgfpatharc{180}{0}{.5\pgf@xc}  
    \pgfpathlineto{\pgfpoint{\pgf@xb}{\pgf@ya}}
    \pgfpathclose
 }
}
\makeatother
\tikzstyle{test1} = [dash circled box, text centered, draw=black, minimum height=1cm,minimum width=1cm]
\tikzstyle{test2} = [barn, text centered, draw=black, minimum height=1cm,minimum width=1cm]
\tikzstyle{gaugedflavor} = [barn,draw, text centered, minimum height=1.2cm,minimum width=1.2cm,draw=black]
\tikzstyle{gaugedflavorS} = [barn,draw, text centered, minimum width=6ex,minimum height=6ex,draw=black]

\pgfdeclareshape{cross}{
  \inheritsavedanchors[from={rectangle}]
    \inheritanchorborder[from={rectangle}] 
     \inheritanchor[from={rectangle}]{north}
    \inheritanchor[from={rectangle}]{center}
    \inheritanchor[from={rectangle}]{south}
    \inheritanchor[from={rectangle}]{west}
    \inheritanchor[from={rectangle}]{east}
\inheritbackgroundpath[from={rectangle}]
    \backgroundpath{
    \draw[solid] (.6ex,-.6ex) -- (-.6ex,.6ex);
    \draw[solid] (.6ex,.6ex) -- (-.6ex,-.6ex);
    }
}

\title{\boldmath Surface operators in ${\mathcal N}=2$ SQCD and Seiberg Duality}
\author[a]{Sujay K. Ashok,}
\affiliation[a]{Institute of Mathematical Sciences \\
Homi Bhabha National Institute (HBNI)\\
IV Cross Road, C.~I.~T.~Campus, \\
  Taramani, Chennai, 600113  Tamil Nadu, India \\}
\emailAdd{sashok@imsc.res.in} 
   
\author[a]{Sourav Ballav,}
\emailAdd{sballav@imsc.res.in}

\author[b,c]{Marialuisa Frau,}
\emailAdd{frau@to.infn.it}
\affiliation[b]{Universit\`a di Torino, Dipartimento di Fisica}

\affiliation[c]{I.\,N.\,F.\,N. - sezione di Torino, \\
Via P. Giuria 1, I-10125 Torino, Italy\\}

\author[b,c]{and Renjan Rajan John\,}
\emailAdd{renjan.rajan@to.infn.it}

\abstract{We study half-BPS surface operators in $\mathcal{N}=2$ supersymmetric asymptotically conformal  gauge theories in four dimensions with SU$(N)$ gauge group and 2$N$ fundamental flavours using localization methods and coupled 2d/4d quiver gauge theories. We show that contours specified by a particular Jeffrey-Kirwan residue prescription in the localization analysis map to particular realizations of the surface operator as flavour defects. 
Seiberg duality of the 2d/4d quivers is mapped to contour deformations of the localization integral which in this case involves a residue at infinity. 
This is reflected as a modified Seiberg duality rule that shifts the Lagrangian of the purported dual theory by non-perturbative terms.
The new rules, that depend on the 4d gauge coupling, lead to a match between the low energy effective twisted chiral superpotentials for any pair of dual 2d/4d quivers. 
}


\begin{document}
\maketitle
\flushbottom

\section{Introduction and Summary}

Surface operators are co-dimension 2 generalisations of 't Hooft and Wilson loops in gauge theories. In this paper, we study surface operators in $\mathcal{N}=2$ SQCD theories with gauge group SU$(N)$ and 2$N$ fundamental flavours in four dimensions. The condition on the number of fundamental flavours ensures that in the limit they are massless the theory is super-conformal at the quantum level. We will refer to these as asymptotically conformal gauge theories. Our interest is in the low-energy effective action of such theories on the Coulomb branch, in the presence of a surface defect. This effective action is encoded in two holomorphic functions: the prepotential, which describes the four dimensional (4d) dynamics without the defect, and the twisted chiral superpotential, which describes the dynamics of the two dimensional (2d) theory on the defect. 

In our study of surface operators we follow two approaches.
In the first approach, we consider the ramified instanton partition function $Z_{\text{inst}}$, which is obtained by a suitable orbifold of the instanton moduli space of the 4d SQCD theory without the defect \cite{Kanno:2011fw} (see also \cite{Ashok:2017odt} for details). One way to realize the instanton moduli space is by considering the open string excitations of D(-1)/D3/D7-brane systems in an orbifold of type IIB string theory. In this realization, the ramified instanton moduli are open strings with at least one end-point on the D(-1)-branes and, using localization techniques, the partition function $Z_{\text{inst}}$ can be written as a contour integral over those moduli which represent the position of the D(-1)-branes in the directions transverse to both the D3 and the D7-branes. For a particular contour whose residues have an interpretation as Young tableaux, this is interpreted as the partition function of a monodromy defect in the gauge theory \cite{Gukov:2006jk,Gukov:2008sn}. Such surface defects are labelled by Levi subgroups of SU$(N)$ which are classified by partitions of $N$. For asymptotically conformal SQCD, it turns out that the flavour group SU$(2N)$ is also broken at the location of the defect into $M$ factors, whose ranks are determined by the same partition of $N$.  Both the prepotential and the twisted chiral superpotential on the Coulomb branch can then be extracted from $Z_{\text{inst}}$ in the limit of vanishing $\Omega$-deformation parameters \cite{Alday:2009fs,Alday:2010vg}. 

In the second approach, we describe surface defects as flavour defects, which are coupled 2d/4d systems realized as quiver gauge theories \cite{Gaiotto:2009fs,Gaiotto:2013sma}. Here the 2d sector is a (2,2) theory described by a gauged linear sigma model in the ultraviolet, in which the vacuum expectation values of the adjoint scalar of the 4d theory act as twisted masses
\cite{Witten:1993yc,Hanany:1997vm}.
The 2d theory has a discrete set of massive vacua determined by solutions of twisted chiral ring equations that extremize the twisted superpotential. Quiver realizations of surface defects can be compared with the localization approach by considering the low energy twisted chiral superpotential on the 4d Coulomb branch.
  
Work along this direction has been pursued in the pure 4d theory in \cite{Ashok:2017odt,Gorsky:2017hro,Ashok:2017bld,Ashok:2017lko,Ashok:2018zxp}. One of the main results in \cite{Ashok:2018zxp} is that there can be different 2d/4d quivers which realize the same flavour defect and are related by 2d Seiberg duality \cite{Benini:2014mia}.
Seiberg duality is an infrared equivalence such that for dual quivers the 
low energy effective superpotentials, evaluated in 
particular vacua, match. These statements are reflected on the localization side in an elegant way: each Seiberg dual realization of the surface operator is associated to a contour prescription and residue theorems guarantee the equality of the low energy effective superpotentials. 
The contours are specified by the Jeffrey-Kirwan (JK) prescription \cite{JK1995} and each reference JK vector associated to a given 2d/4d quiver can be written unambiguously in terms of its Fayet-Iliopoulos (FI) parameters. 

The localization integrand in the asymptotically conformal case differs from that of the pure theory only in the structure of the numerator and hence the set of poles picked by a given JK vector remains the same as in the theory without flavours. As a result, on the quiver side, the ranks of gauge nodes in the quiver remain the same. The ranks of flavour nodes are uniquely fixed by how the flavour symmetry is broken by the defect and by requiring conformality at each 2d gauge node. For each contour choice, we propose how to construct a 2d/4d quiver theory whose twisted superpotential, when evaluated on the solutions of the twisted chiral ring equations, matches the localization result after a suitable map of parameters. Some work in this direction appeared recently in \cite{Baek:2018vdw}, but our analysis of Seiberg duality has significant differences.

As mentioned earlier, in the case of the pure 4d gauge theory, distinct contour choices are equivalent and give rise to Seiberg-dual 2d/4d gauge theories. However, there is a new feature in the asymptotically conformal SQCD case: due to a non-vanishing residue at infinity, distinct contours are inequivalent. While the prepotential obtained from the instanton partition function is independent of the contour of integration, the twisted superpotential turns out to be different for distinct contour choices.  

The main focus of this work is to understand how Seiberg duality can be consistent with such inequivalent contours in the context of surface defects in asymptotically conformal SQCD. The ranks and connectivity of the quivers one gets by Seiberg duality are exactly those that correspond to the different JK prescriptions,
but the effective twisted superpotentials on the 4d Coulomb branch for different JK prescriptions are not trivially related.
The resolution to this is known for the case of 2d gauge theories in which the flavour group is not gauged \cite{Benini:2014mia}: the Lagrangian of the dual theory is modified by non-perturbative corrections. Our main result in this work is a proposal for a generalized Seiberg duality rule 
with further non-perturbative terms that applies to the case of surface operators realized as flavour defects.
We derive this from the localization integrand by a careful analysis of the residue at infinity. With the modified duality rules that now also involve the 4d gauge coupling, the twisted superpotentials evaluated on the solutions of the chiral ring equations match for all dual pairs of theories. 

The rest of the paper is organized as follows. In Section 2, we introduce surface operators as monodromy defects and write the localization integrand from which the instanton partition function is obtained after specifying a contour of integration. In Section 3, we relate the different contours of integration to distinct 2d/4d quivers by studying the $[p,N-p]$ defect. In Section 4, we propose a generalized Seiberg duality move and show in the case of the simplest quiver in what manner the Lagrangian for the dual quiver is corrected from the perturbatively exact 1-loop result by non-perturbative terms. In Section 5, we analyze the 3-node quiver in detail and test successfully the rules laid out in Section 4. We provide a derivation of our proposal using localization methods in Appendix A and collect some details of the computations in the remaining Appendices. 

\section{Surface operators as monodromy defects}

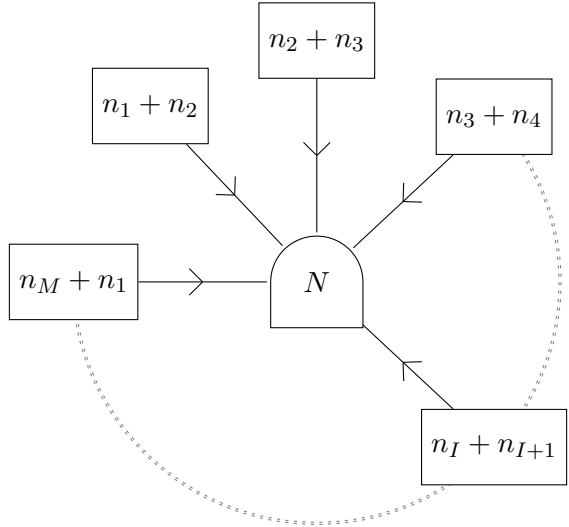
\begin{figure}[H]
\begin{center}
\begin{tikzpicture}[decoration={
markings,
mark=at position 0.5 with {\draw (-5pt,-5pt) -- (0pt,0pt);
                \draw (-5pt,5pt) -- (0pt,0pt);}}]
\def\ra{3.2cm};
\node(Ncopy)[circle, minimum height=1.3cm,minimum width=1cm]{};
\draw[white,text=black] ++(-\ra,0) arc(180:540:\ra)  node(nM)[pos=0, squashedflavor,fill=white]{$n_M+n_1$} node(n1)[pos=0.87, squashedflavor]{$n_1+n_2$} node(n2)[pos=0.75, squashedflavor]{$n_2+n_3$} node(n3)[pos=0.62, squashedflavor,fill=white]{$n_3+n_4$} node(nI)[pos=0.38, squashedflavor,fill=white]{$n_I+n_{I+1}$};
\graph[edges={postaction={decorate}}]{ (nM)--(Ncopy); (n1)--(Ncopy);(n2)--(Ncopy);(n3)--(Ncopy);(nI)--(Ncopy);};
\draw[text=black] (Ncopy) node(N)[gaugedflavor,fill=white]{$N$};
\begin{scope}[on background layer]
\draw[dotted,gray, very thick] (-\ra,0) arc(180:405:\ra) ;
\end{scope}
\end{tikzpicture}
\end{center}
\vspace{-0.5cm}
\caption{The asymptotically conformal 4d node with broken flavour symmetry. In the realization of surface operators as flavour defects, the 4d gauge node as well as the 4d flavour nodes act as matter for the gauge nodes of the 2d quiver. }
\label{conformal4dnode}
\end{figure}

The 4d ${\mathcal N}=2$ gauge theory of interest is the asymptotically conformal  SQCD, which is an SU$(N)$ gauge theory with $2N$ fundamental flavours. We are interested in half-BPS surface operators in this gauge theory, whose classification is the same as that for the pure gauge theory studied in \cite{Ashok:2018zxp}. For every partition of $N$, given by
\be
\sum_{I=1}^M n_I = N\,,
\ee
one obtains a surface operator, labelled by the Levi subgroup:
\be
\mathbb{L} = S\left[U(n_1)\times U(n_2)\times\ldots \times U(n_M)\right] \,.
\ee
The Coulomb vevs $a_u$ (with $u=1,\ldots, N$ and such that $\sum_u a_u=0$) naturally split into $M$ sets, each of which contains $n_I$ elements. Without loss of generality, we can order them from $1$ to $N$ and make the $M$ partitions sequentially:
\begin{equation}
\label{asplit}
\big\{a_1,\ldots,  a_{r_1}|\ldots \big|
a_{r_{I-1}+1}, \ldots, a_{r_{I}}\big|
\ldots |a_{r_{M-1} +1}, \ldots, a_N \big\}
\end{equation}
where we have defined integers $r_I$ such that
\begin{align}
\label{rI}
r_I = \sum_{J=1}^{I} n_J \,. 
\end{align}
To avoid cumbersome expressions, we introduce a convenient notation, and for $I=1, \ldots M$, we define 
\begin{align}
{\cal N}_I \equiv \{ r_{I-1}+1,  r_{I-1}+2,\ldots, r_I \} \,,
\end{align}
which is a set of cardinality $n_I$. Here and in the following, we use the convention that the index $I$ is periodic 
mod $M$.
A new feature of surface operators in the asymptotically conformal SQCD theory is that it also breaks the SU$(2N)$ flavour symmetry to the following subgroup:
\be\label{flavourbreaking}
\mathbb{F} = S\left[U(n_1+n_2)\times U(n_2+n_3)\times \ldots \times U(n_M + n_1) \right]\,.
\ee
To denote the blocks into which the flavour group is broken, it is useful to define 
\be
{\mathcal F}_I = \{ r_{I-1}+r_I - r_1 +1, \ldots , r_I + r_{I+1}-r_1\} \,,
\ee
which is a set of cardinality $n_I+n_{I+1}$.
The breaking of flavour symmetry in the presence of the surface operator is represented in Fig. \ref{conformal4dnode}.

\noindent
The instanton partition function in the presence of such a surface operator, which is also referred to as the ramified instanton partition function, can be derived from the moduli action of a D(-1)/D3/D7-brane system 
in an orbifold background that represents the surface defect. Given the breaking of the gauge and flavour symmetry groups, the analysis is very similar to what was carried out in \cite{Ashok:2017odt} and therefore here we merely present the answer: 
\begin{equation}
Z_{\text{inst}}[\vec{n}] = \sum_{\{d_I\}}Z_{\{d_I\}}[\vec{n}]\quad
\mbox{with}~~~Z_{\{d_I\}}[\vec{n}]= \prod_{I=1}^M \Big[\frac{(-q_I)^{d_I}}{d_I!}
\int \prod_{\sigma=1}^{d_I} \frac{d\chi_{I,\sigma}}{2\pi\ii}\Big]~
z_{\{d_I\}}
\label{Zso4d5d}
\end{equation}
where
\begin{align}
z_{\{d_I\}} & = \,\prod_{I=1}^M \prod_{\sigma,\tau=1}^{d_I}\,
\frac{\left(\chi_{I,\sigma} - \chi_{I,\tau} + \delta_{\sigma,\tau}\right)
}{\left(\chi_{I,\sigma} - \chi_{I,\tau} + \epsilon_1\right)}
\times\prod_{I=1}^M \prod_{\sigma=1}^{d_I}\prod_{\rho=1}^{d_{I+1}}\,
\frac{\left(\chi_{I,\sigma} - \chi_{I+1,\rho} + \epsilon_1 + \hat\epsilon_2\right)}
{\left(\chi_{I,\sigma} - \chi_{I+1,\rho} + \hat\epsilon_2\right)}
\label{zexplicit5d}
\\
& ~~\times
\prod_{I=1}^M \prod_{\sigma=1}^{d_I} \frac{\prod_{i\in {\mathcal F}_I} (\chi_{I,\sigma} - m_{i})}
{\prod_{s\in{\mathcal N}_I}\left(a_{s}-\chi_{I,\sigma} + \frac 12 (\epsilon_1 + \hat\epsilon_2)\right)
\prod_{t\in{\mathcal N}_{I+1}}
\left(\chi_{I,\sigma} - a_{t} + \frac 12 (\epsilon_1 + \hat\epsilon_2)\right)}~.
\nonumber
\end{align}
The $M$ positive integers $d_I$ count the numbers of ramified instantons in the various sectors
and $\epsilon_1$ and $\hat\epsilon_2=\frac{\epsilon_2}{M}$ parametrize the $\Omega$ background introduced to localize the integration over the instanton moduli space \cite{Nekrasov:2002qd,Nekrasov:2003rj}. 
If one neglects the contribution of the flavours, namely the numerator factors in the second line of \eqref{zexplicit5d}, the integrand is identical to that of the pure 4d theory. Note that the flavour factors are such that the breaking of the flavour symmetry is respected. 
Unlike the case of surface operators in the pure theory, in asymptotically conformal SQCD, the instanton counting parameters $q_I$ are dimensionless. 

The low energy effective action that governs the combined gauge theory/surface operator system is completely specified by two holomorphic functions, the prepotential ${\mathcal F}$ and the twisted chiral superpotential ${\mathcal W}$. The non-perturbative contributions to these functions are obtained by taking the vanishing limit of the $\Omega$-deformation parameters:
\be\label{Zepszero}
\lim_{\epsilon_i\rightarrow 0}\log \left(1+Z_{\text{inst}}[\vec{n}]\right) = -\frac{{\mathcal F}_{\text{inst}}}{\epsilon_1\epsilon_2}+\frac{{\mathcal W}_{\text{inst}}}{\epsilon_1}~.
\ee
The physical interpretation of ${\mathcal W}_{\text{inst}}$ for a generic partition can be given by studying coupled 2d/4d quiver theories \cite{Gaiotto:2013sma, Ashok:2017lko}. 
In the following section we briefly review the relation between the localization analysis and the 2d/4d field theory interpretation of surface operators, focusing in particular on the connection between the choice of the integration contour in the localization integrals and the choice of a representative 2d quiver theory in a sequence of Seiberg dual models.

\section{Contours and quivers}

In order to extract the twisted chiral superpotential from the localization analysis one needs to provide a residue prescription to calculate the instanton partition function. This
prescription is most succinctly specified via a JK reference vector \cite{JK1995} 
that uniquely specifies the set of poles chosen by the contour %
\footnote{For applications to gauge theories see, for instance, \cite{Benini:2013nda, Hori:2014tda, Gorsky:2017hro}.}.

In our previous work \cite{Ashok:2018zxp}  on surface operators in the pure 4d theory, it was shown that  different JK prescriptions map to distinct 2d/4d quiver gauge theories.
In that case the quivers are equivalent, and related to each other by Seiberg duality. For such quivers the ranks of the 2d gauge groups directly correlate with a choice of a massive vacuum and the evaluation of the twisted chiral superpotential in that particular massive vacuum reproduces the twisted superpotential ${\mathcal W}_{\text{inst}}$ calculated using localization. Despite the equivalence between quivers, we could obtain an unambiguous map between contours and quivers, thanks to the match between the individual residues on the localization side and the individual terms in the solution of the chiral ring equations. In particular, the number of residues that contribute to the contour integrals are related to the ranks of the 2d nodes of the quivers, while the coefficients of the JK vector correspond to the FI parameters of the 2d gauge groups.

In extending this correspondence to the asymptotically conformal SQCD case, we have to consider the following relevant points:

\begin{enumerate}
\item 
If we consider the form of $Z_{\text{inst}}$ given in \eqref{zexplicit5d} one observes that, for a given contour prescription, the set of poles that contribute to the localization integral is identical to those that contribute in the pure gauge theory. This is obvious given that the fundamental flavours only add factors in the numerator of the instanton partition function, leaving the denominator and its singularity structure unchanged. 
It therefore follows that the quiver we may associate to a given integration contour has the same 2d gauge content of the one in the corresponding case without flavour. In particular the ranks of the 2d gauge nodes remain identical.

\item The main difference with the pure case is that the matter multiplets now
provide flavours to the 2d gauge nodes as well. 
The additional constraint we have to impose is that for every quiver the complexified FI parameters of the 2d gauge nodes should not run, 
so that the 2d gauge theories are conformal. Since the ranks of the 2d gauge nodes are fixed, the number of (anti-) fundamental flavours at each node is fixed by the necessity to cancel the contribution of the neighbouring gauge nodes (that also act as flavours) to the running of the FI coupling. Combined with the breaking of the flavour symmetry to $\mathbb{F}$ as in \eqref{flavourbreaking} for a generic surface operator, this uniquely fixes how the broken 4d flavour group acts on the 2d gauge nodes. 

\item Given the resulting 2d/4d quiver, the evaluation of the twisted chiral superpotential in the chosen vacuum once again reproduces the twisted superpotential ${\mathcal W}_{\text{inst}}$ calculated using localization and with the particular JK prescription associated to the 2d/4d quiver.

\end{enumerate}
We illustrate the points above for a generic surface operator $[n_1, n_2, \ldots n_M]$ by giving a particular 2d/4d realization, shown in Figure \ref{genericsurfaceop} below.

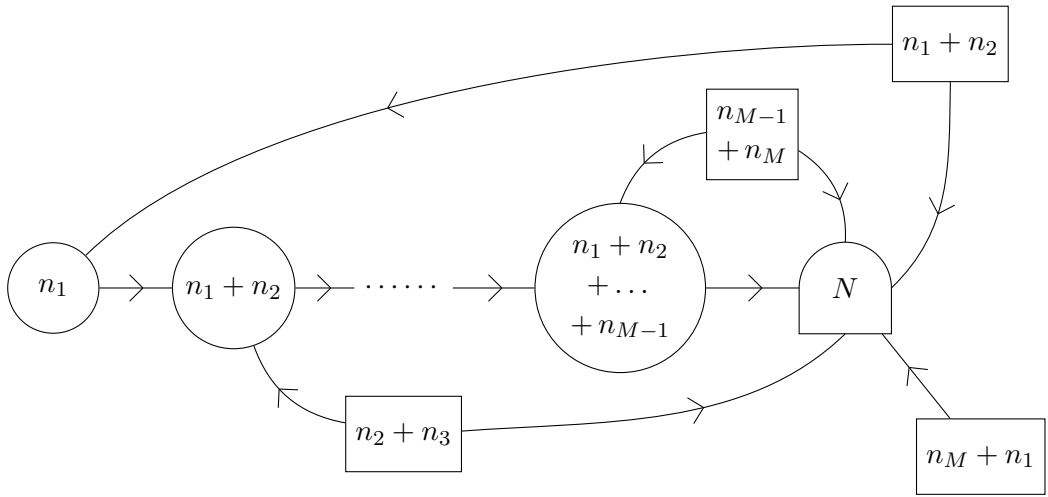
\begin{figure}[H]
\begin{tikzpicture}[decoration={
markings,
mark=at position 0.6 with {\draw (-5pt,-5pt) -- (0pt,0pt);
                \draw (-5pt,5pt) -- (0pt,0pt);}}]
  \matrix[row sep=3mm,column sep=0mm] {
     &\node{$\phantom{n_3+}$}; & & \node{$\phantom{n_3}$}; &&\node{$\phantom{n_3+}$};&&\node[](AfNW)[squashedflavor,align=center,minimum height=1.2cm] {$n_{M-1}$\\$+\, n_M$};  &&\node[xshift=-4mm,yshift=+12mm](AfNE)[squashedflavor] {$n_1+n_2$}; \\      \node(Ag1)[gauge] {$n_1$};  & & \node(Ag2)[gauge] {$n_1+n_2$}; & & \node(Adots){$\cdots\cdots$}; && \node(AgM)[gauge,align=center] {$n_1 + n_2$\\$+\dots$\\$+\, n_{M-1}$}; & &\node(Agf)[gaugedflavor]{$N$};&\\
      & & &&\node(AfSW)[squashedflavor] {$n_2+n_3$};  & &&&&\node[yshift=-3mm](AfSE)[squashedflavor] {$n_M+n_1$}; \\
   };
\graph[edges={postaction={decorate}}]{
(Ag1) --(Ag2)--(Adots)--(AgM)--(Agf);
(AfNE)--[bend right,looseness=0.7,out=-15](Ag1);(AfNE)--[bend left,right anchor=east,left anchor=south,out=15](Agf);
(AfNW)--[bend right,right anchor=north](AgM);(AfNW)--[bend left,right anchor=north](Agf);
(AfSW)--[bend left](Ag2);(AfSW)--[bend right,right anchor=south,out=-10,looseness=.9](Agf);
(AfSE)--(Agf);
};
\end{tikzpicture}
\caption{One realization of the $[n_1, n_2, \ldots n_M]$ surface operator as a 2d/4d quiver in which the 2d gauge nodes are oriented. The 4d flavour groups provide matter for the 2d nodes such that the $\beta$-function of each FI parameter is zero.}
\label{genericsurfaceop}
\end{figure}

One property we would like to emphasize is that the map between a given JK contour and its corresponding quiver is unambiguous: the superpotential calculated from a particular localization prescription and that obtained from the twisted chiral ring equations of the 2d/4d quiver match. We will show this explicitly in the following sections in various examples. 

It is important to stress that in discussing points $1, 2,$ and  $3$ above, nowhere did we use Seiberg duality rules for the asymptotically conformal SQCD theory or mention the equivalence between the different 2d/4d coupled theories. For surface operators in the pure 4d theory, the 2d/4d quivers related to distinct JK prescriptions were part of a duality chain in which each step is a particular 2d Seiberg duality move (for example, see Fig.~7 in \cite{Ashok:2018zxp}). In that case it was true that the twisted superpotential calculated using different contour choices were identical. However in asymptotically conformal SQCD there is an interesting twist to the story. By explicit calculation, one can check that, while the prepotential calculated using the above prescription is independent of the choice of surface operator and of the contour of integration, the twisted chiral superpotential calculated using different JK prescriptions, in fact, do {\it not} agree.  We illustrate this point in the simple setting of the $[p,N-p]$ defect.

\subsection{The $[p,N-p]$ defect} 
Let us consider the surface defect $[p,N-p]$, and focus for simplicity on the 1-instanton contribution to the partition function $Z_{1\text{-inst}}$.

In order to write it in compact form, we introduce the polynomials:
\begin{align}
\label{PandBsplits}
P_I(z) &= \prod_{u\in {\mathcal N}_I}(z-a_u)\,, \qquad
B_I(z) = \prod_{i\in {\mathcal F}_I}(z-m_i)\,.
\end{align}
In terms of these, $Z_{1\text{-inst}}$ for the 2-node defect takes the form
\begin{equation}
Z_{1\text{-inst}} = -\sum_{I=1}^2\, \frac{q_I}{\epsilon_1}\!
 \int \!\frac{d\chi_I}{2\pi \ii}\, 
\frac{(-1)^{n_I}\,B_I(\chi_I)}{P_I\left(\chi_I - \frac 12 (\epsilon_1 + \hat\epsilon_2)\right)P_{I+1}\left(\chi_I 
+ \frac 12 (\epsilon_1 + \hat\epsilon_2)\right)} \,,
\label{Z1inst}
\end{equation}
while, using (\ref{Zepszero}), the 1-instanton twisted superpotential is given by
\begin{equation}
\cW_{1\text{-inst}} =\lim_{\epsilon_i\,\rightarrow 0}\epsilon_1  Z_{1\text{-inst}} \,.
\end{equation}
To compute the integrals in \eqref{Z1inst} we can use distinct JK prescriptions that 
simply correspond to integrating each $\chi_I$ along a closed contour in the upper $(+)$ or lower $(-)$ half-planes. 
According to the analysis of \cite{Ashok:2017lko,Ashok:2018zxp}, out of the four inequivalent possibilities, only two JK prescriptions 
are relevant and we denote them by $(+-)$ and $(-+)$, respectively.

With the $(+-)$ prescription, the 1-instanton contribution to the twisted superpotential is
\begin{align}
\label{LOCJK1}
{\mathcal W}^{+-}_{1\text{-inst}}&=(-1)^{p+1}q_1\sum_{u\in\mathcal{N}_1}\frac{B_1(a_u)}{P'_1(a_u)P_2(a_u)}
+(-1)^{N-p}q_2\sum_{u\in\mathcal{N}_1}\frac{B_2(a_u)}{P'_1(a_u)P_2(a_u)}\,,
\end{align}
while with the $(-+)$ prescription we get
\begin{align}
\label{LOCJK2}
{\mathcal W}^{-+}_{1\text{-inst}}&=(-1)^{p}q_1\sum_{u\in\mathcal{N}_2}\frac{B_1(a_u)}{P_1(a_u)P'_2(a_u)}
+(-1)^{N-p+1}q_2\sum_{u\in\mathcal{N}_2}\frac{B_2(a_u)}{P_1(a_u)P'_2(a_u)}\,.
\end{align}
where the $\prime$ symbol denotes derivative.
%
We can easily verify that the two superpotentials
are different and that their difference is
\be
\label{deltaW1}
 {\mathcal W}^{-+}_{1\text{-inst}}- {\mathcal W}^{+-}_{1\text{-inst}}  = (-1)^{p} q_1 \sum_{u\in\mathcal{N}_1\cup\mathcal{N}_2} \frac{B_1(a_u)}{P'(a_u)} 
+(-1)^{N-p+1} q_2\sum_{u\in\mathcal{N}_1\cup\mathcal{N}_2} \frac{B_2(a_u)}{P'(a_u)} \,,
\ee
where $P(z)$ is the classical gauge polynomial given by 
\begin{align}
\label{CGP}
P(z)=\prod_{u=1}^N\,(z-a_u)\,. 
\end{align}

It is simple to realize that, since the difference of the contours in the upper and lower half-planes is a contour around infinity, the difference
\eqref{deltaW1} is due to non-vanishing residues at infinity in the integrand of $Z_{1\text{-inst}}$, a property which is characteristic of 
asymptotically conformal theories. So we can write:
\begin{align}
\label{JK1JK2difference}
 {\mathcal W}^{-+}_{1\text{-inst}}- {\mathcal W}^{+-}_{1\text{-inst}} 
&= \int_{C_\infty} dz \left[ (-1)^{p} q_1  \frac{B_1(z)}{P(z)} +(-1)^{N-p+1}q_2 \frac{B_2(z)}{P(z)}  \right]\notag\\[3mm]
&= \Big[(-1)^{p+1} q_1+(-1)^{N-p+1}q_2\Big]  \sum_{i\in {\mathcal F}_1} m_i\,
\end{align}
where $C_\infty$ is a closed curve encircling infinity clockwise, and the second line follows from using the explicit expressions  for the gauge and flavour polynomials.

In Appendix \ref{contourdeform}, by lifting the model to five dimensions with one compact direction, we show that the existence of non-vanishing residues at infinity holds at every instanton number. This explains why twisted superpotentials evaluated with different contour prescriptions are generically different. For the 2-node defect $[p, N-p]$, we are able to resum the instanton expansion and obtain
\be\label{DeltaWexact}
{\mathcal W}^{-+}_{\text{inst}} - {\mathcal W}^{+-}_{\text{inst}} = -\Big[\log(1+(-1)^{p}q_1) +\log(1+(-1)^{N-p}q_2) \Big] \sum_{i\in {\mathcal F}_1} m_i\,.
\ee

As discussed earlier, we expect that different JK prescriptions map to distinct quivers related 
by 2d Seiberg duality, with equivalent superpotentials.
This result therefore suggests that in SQCD with surface defects, 
the definition of what is the dual quiver necessarily involves non-perturbative modifications due to ramified
instantons. We will discuss this issue in greater detail in the following section.  

\section{Generalized Seiberg duality} 
\label{GSD}

\begin{figure}[H]
\begin{centering}
{\footnotesize
\begin{tikzpicture}[decoration={
markings,
mark=at position 0.6 with {\draw (-5pt,-5pt) -- (0pt,0pt);
                \draw (-5pt,5pt) -- (0pt,0pt);}}]
  \matrix[row sep=10mm,column sep=5mm, ampersand replacement=\&] {
      \node(g1)[flavor] {$N_f$};  \& \& \node(g2)[gauge] {$N$}; 
      \& \& \node(g3)[flavor]{$N_f$};\\
  };
\graph{(g1) --[postaction={decorate}](g2)--[postaction={decorate}](g3);};
\end{tikzpicture}
}
\caption{The gauge group is represented by a circle, and the flavour groups are represented by squares. The quiver diagram has a single 2d gauge node of rank $N$ with $N_f$ fundamental and $N_f$ anti-fundamental flavours attached to it.}
\label{PreBasicDuality}
\end{centering}
\end{figure}
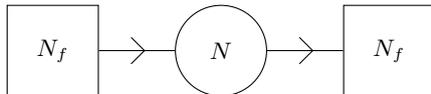

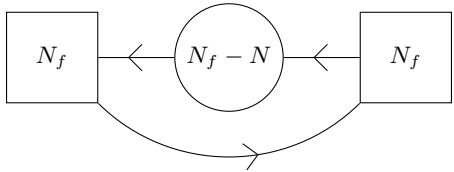
\begin{figure}[H]
\begin{centering}
{\footnotesize
\begin{tikzpicture}[decoration={
markings,
mark=at position 0.6 with {\draw (-5pt,-5pt) -- (0pt,0pt);
                \draw (-5pt,5pt) -- (0pt,0pt);}}]
  \matrix[row sep=10mm,column sep=5mm,ampersand replacement=\&] {
      \node(g1)[flavor] {$N_f$};  \& \& \node(g2)[gauge] {$N_f-N$}; 
      \& \&\node(g3)[flavor]{$N_f$};\\
  };
\graph{(g2) --[postaction={decorate}](g1) (g3)--[postaction={decorate}](g2);};
\graph{(g1) --[out=-45,in=-135,postaction={decorate}](g3)};
\end{tikzpicture}
}
\caption{The quiver diagram obtained after a 2d Seiberg duality on the gauge node in Fig.~\ref{PreBasicDuality}.}
\label{PostBasicDuality}
\end{centering}
\end{figure}

We now study Seiberg duality in the 2d/4d quiver realization of the defect and propose a relation between the twisted superpotentials of dual quivers. For the purely 2d case, this has been discussed in detail in \cite{Benini:2014mia}. We begin with a 2d U($N$) gauge theory with $N_f$ fundamental flavours and $N_f$ anti-fundamental flavours shown in Figure~\ref{PreBasicDuality}.
\noindent
We now perform a Seiberg duality operation on the 2d gauge node, and obtain the quiver diagram shown in Figure~\ref{PostBasicDuality}. 
\noindent
Under the duality, the roles of the fundamental and anti-fundamental flavours are exchanged as denoted by the reversal 
of the arrows. There is also the addition of a mesonic field, as described by the line connecting the two flavour groups. 

When these duality rules are applied to quiver theories, one has to take into account that for each 2d gauge node flavours can be provided by 
other 2d gauge nodes of the quiver;
in such cases the extra mesonic field should be treated as just another chiral multiplet in the dual quiver. We will see several examples in later sections. So far, we have only shown how the quiver itself is modified by the action of duality,
we still have to show how the duality acts on the parameters and Lagrangian of the quiver theories. 

We again focus on the simplest 2-node case and solve the twisted chiral ring equations of the two purported dual quivers. Imposing duality will then allow us to find the rules. 

\subsection{The 2-node case}
We consider the $[p, N-p]$ defect and describe its effective action. We remark that the results of this subsection have some partial overlap with those of the recent paper \cite{Baek:2018vdw}, but our analysis of Seiberg duality has significant differences. 

\begin{figure}[H]
\begin{center}
{\footnotesize
\begin{tikzpicture}[decoration={
markings,
mark=at position 0.6 with {\draw (-5pt,-5pt) -- (0pt,0pt);
                \draw (-5pt,5pt) -- (0pt,0pt);}}]
  \matrix[row sep=10mm,column sep=10mm] {
      &\node(Nup)[flavor] {\Large{$N$}};  \\ 
      \node(g1)[gauge] {\Large{$p$}}; & \node(gfN)[gaugedflavor] {\Large{$N$}};\\
      &\node(Ndown)[flavor] {\Large{$N$}};  \\ 
  };
\graph{(Nup) --[postaction={decorate}](gfN) (Ndown) --[postaction={decorate}](gfN)};
\graph{(Nup) --[out=180,in=90,postaction={decorate}](g1) --[postaction={decorate}](gfN)};
\end{tikzpicture}
}
\end{center}
\caption{A 2d/4d quiver realization of the $[p,N-p]$ defect in SU$(N)$ theory with $2N$ 
flavours}
\label{Q1pN-p}
\end{figure}
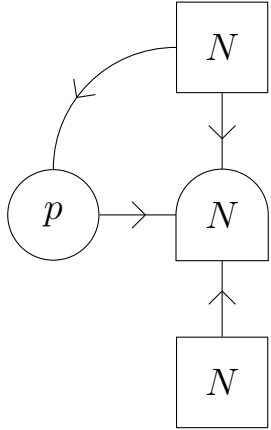

\paragraph{The quiver $Q_0$:} We first consider the realization of the defect as the quiver in Figure \ref{Q1pN-p}.
After the massive chiral multiplets are integrated out, the twisted chiral superpotential takes the following form:
\be
\label{WpN-p}
{\mathcal W}_{Q_0} = \log x\, \sum_{s\in\mathcal{N}_1} \sigma_s - \sum_{s\in\mathcal{N}_1}\sum_{i\in\mathcal{F}_1} \varpi(m_i -\sigma_s) - \sum_{s\in\mathcal{N}_1} \Big\langle \Tr\varpi(\sigma_s-\Phi)\Big\rangle\,,
\ee
where $x$ is the exponentiated FI parameter of the 2d theory, $\sigma_s$ are the scalars in the twisted chiral suprerfield that encodes the 2d vector multiplet and the $m_i$ are the masses of the 4d flavours that also act as twisted masses for the 2d chiral multiplets. We have also introduced the function $\varpi(x) = x\big(\log \frac{x}{\mu} - 1\big)$, which is the result of integrating out a chiral multiplet of twisted mass $x$. In the last term of \eqref{WpN-p}, the angular brackets
denote a chiral correlator in the 4d SU($N$) gauge theory, and $\Phi$ is the adjoint scalar in the vector multiplet.  The twisted chiral ring equations are
\cite{Nekrasov:2009ui,Nekrasov:2009rc}:
\begin{align}
\exp\,\left(\frac{\partial {\mathcal W_{Q_0}}}{\partial\sigma_s}\right)=1,\quad s\in\mathcal{N}_1\,,
\end{align}
and they explicitly read
\begin{align}
\label{2nodeQ1TCR1crude}
\exp\Big\langle \Tr\log(\sigma_s-\Phi)\Big\rangle = (-1)^N\,x\,B_1(\sigma_s)\quad\text{for}\ s\in\mathcal{N}_1
\end{align}
where $B_1(\sigma_s)$ is the polynomial defined in \eqref{PandBsplits}.
For the asymptotically conformal case under consideration, the resolvent of the 4d gauge theory which determines the chiral correlator, has a non-trivial dependence on $q_0$, the instanton weight of the 4d SU$(N)$ theory. We refer the reader to Appendix \ref{resolvent} for details; here we merely present the result, namely
\begin{align}
\label{resolventfromapp}
\left\langle \Tr \log \frac{z-\Phi}{\mu} \right\rangle = \log\bigg((1+q_0)\,\frac{\widehat P(z)+Y}{2\mu^N}\bigg) \,.
\end{align}
Here, $\widehat P(z)$ is the quantum gauge polynomial corresponding to the classical 
one defined in \eqref{CGP}:
\be
\widehat P(z) = z^N + u_2 z^{N-2} + \ldots + (-1)^Nu_N \,,
\ee
where $u_k$ are the gauge invariant coordinates on moduli space, and the variable
$Y$ is given in terms of the Seiberg-Witten curve of the asymptotically conformal 4d gauge theory:
\be
\label{SWcurve}
Y^2 =\widehat P(z)^2 - \frac{4q_0}{(1+q_0)^2} B(z) \,,
\ee
where $B(z)$ is the flavour polynomial.
We refer the reader to  Appendix \ref{resolvent} for details.

\begin{figure}[H]
\begin{center}
{\footnotesize
\begin{tikzpicture}[decoration={
markings,
mark=at position 0.6 with {\draw (-5pt,-5pt) -- (0pt,0pt);
                \draw (-5pt,5pt) -- (0pt,0pt);}}]
  \matrix[row sep=10mm,column sep=10mm] {
      &\node(Nup)[flavor] {\Large{$N$}};  \\ 
      \node(g1)[gauge] {\Large{$N-p$}}; & \node(gfN)[gaugedflavor]{\Large{$N$}};\\
      &\node(Ndown)[flavor] {\Large{$N$}};  \\ 
  };
\graph{(Nup) --[postaction={decorate}](gfN) (Ndown) --[postaction={decorate}](gfN)};
 \graph{(g1) --[out=90,in=180,postaction={decorate}](Nup) (gfN) --[postaction={decorate}](g1)};
\end{tikzpicture}
}
\end{center}
\caption{The 2d/4d quiver diagram obtained after the action of Seiberg duality on the 2d node in Fig.~\ref{Q1pN-p}}
\label{Q2pN-p}
\end{figure}
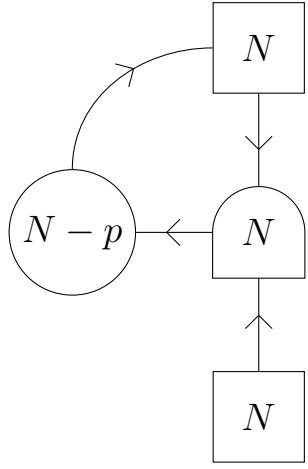

Exponentiating \eqref{resolventfromapp} and using \eqref{SWcurve},  we can recast the twisted chiral ring equations \eqref{2nodeQ1TCR1crude} in the following form
\begin{align}
\label{2nodeQ1TCR1}
(1+q_0)\,\,\widehat P(\sigma_s) \,&= (-1)^{N}\,\Bigg( x\,B_1(\sigma_s)\,\, + \,\,\frac{q_0 }{x} \, B_2(\sigma_s)\Bigg)\quad\text{for}\ s\in\mathcal{N}_1.
\end{align}
The classical vacuum about which we solve these equations is
\be
\sigma_s = a_s +\delta\sigma_s \quad\text{for}\quad s\in\mathcal{N}_1\,,
\ee
and the solution in the 1-instanton approximation is
\begin{align}
\delta\sigma_s &=(-1)^{N}\frac{1}{P'_1(a_s)P_2(a_s)}\left[\,x\, B_1(a_s)\,+\,\frac{q_0}{x}\,B_2(a_s)\right]\quad\text{for}\,s\in\mathcal{N}_1\,.
\end{align}
Let us now evaluate the twisted superpotential \eqref{WpN-p} on this solution. This is a little tricky since one needs to expand the 4d chiral correlator 
$\Big\langle \Tr \varpi (\sigma-\Phi )\Big\rangle$ in powers of $q_0$. This is carried out in Appendix \ref{resolvent}; using those results, we find  (neglecting the 1-loop contributions)
\begin{align}
{\mathcal W}_{Q_0}(\sigma_{\star}) =& \log x\, \sum_{s\in{\mathcal N}_1} a_s 
+(-1)^{N}x\sum_{s\in\mathcal{N}_1}\frac{B_1(a_s)}{P'_1(a_s)P_2(a_s)}
+(-1)^{N+1}\frac{q_0}{x}\sum_{s\in\mathcal{N}_1}\frac{B_2(a_s)}{P'_1(a_s)P_2(a_s)}\,\,.
\end{align}
It can be easily checked that the 1-instanton terms match
the localization result \eqref{LOCJK1} with the $(+-)$ prescription, namely
\begin{align}
\mathcal{W}_{Q_0}(\sigma_\star)\big|_{1\text{-inst}}=\mathcal{W}^{+-}_{1\text{-inst}}\, ,
\end{align}
provided we make the following identifications:
\begin{align}
\label{Q0map}
q_1=(-1)^{N+p+1}\,x\,,\quad q_2=(-1)^{p+1}\,\frac{q_0}{x}\,.
 \end{align}
We have checked that the match between the superpotential evaluated on the solution of twisted chiral ring equations and the localization results continues to hold up to 8 instantons for various low rank cases. 
\vspace{.5cm} 
\paragraph{The quiver $Q_1$:} Acting with the Seiberg duality rules on the quiver diagram of Fig.~\ref{Q1pN-p}, one obtains the quiver diagram represented in Fig.~\ref{Q2pN-p}.
After integrating out the massive chiral multiplets, the twisted chiral superpotential 
corresponding to this quiver diagram
takes the following form:
\begin{align}
\label{pN-pQ2W}
{\mathcal W}_{Q_1}&=\log y\,\sum_{s\in \mathcal{N}_2}\,\sigma_s-\sum_{s\in \mathcal{N}_2}
\sum_{i\in \mathcal{F}_1}  \varpi(\sigma_s-m_i)-\sum_{s\in \mathcal{N}_2}\Big\langle\text{Tr}\,\varpi(\Phi-\sigma_s)\Big\rangle\, ,
\end{align}
where we have denoted by $y$ the exponentiated FI parameter 
of of the 2d gauge node.
The chiral ring equations that follow from ${\mathcal W}_{Q_1}$ are:
\begin{align}
\label{2nodeQ2TCR1}
(1+q_0)\,\,\widehat P(\sigma_s) \,&= (-1)^{N}\,\bigg[\frac{1}{y}\,B_1(\sigma_s)\,\, + \,\,q_0\,y\,\,B_2(\sigma_s)\bigg]\quad\text{for}\ s\in\mathcal{N}_2\,.
\end{align}
We solve them in the vacuum given by
\begin{align}
\sigma_s &= a_{s} + \delta\sigma_s \quad\text{for}\quad s\in \mathcal{N}_2 \,,
\end{align}
and in the 1-instanton approximation we obtain
\begin{align}
\delta\sigma_s &=(-1)^{N}\frac{1}{P_1(a_s)P'_2(a_s)}\left[\frac{1}{y}\, B_1(a_s)\,+\,q_0\,y\,B_2(a_s)\right]\quad\text{for}\,\,s\in\mathcal{N}_2\,.
\end{align}
Evaluating the twisted chiral superpotential on this solution, we find (neglecting the 1-loop contributions as before)
\begin{align}
{\mathcal W}_{Q_1}(\sigma_{\star}) = & \log y\, \sum_{s\in{\mathcal N}_2} a_s
+(-1)^{N+1}\frac{1}{y}\sum_{s\in\mathcal{N}_2}\frac{B_1(a_s)}{P_1(a_s)P'_2(a_s)}
+(-1)^{N}q_0\,y\sum_{s\in\mathcal{N}_2}\frac{B_2(a_s)}{P_1(a_s)P'_2(a_s)}\,. 
\end{align}
If we now impose that the classical contributions in ${\mathcal W}_{Q_0}$ and ${\mathcal W}_{Q_1}$ match, we find that the FI parameters of the pair of dual theories 
are related in the same way as in the 
pure theory, namely:
\be
\label{yvsx2node}
y= \frac{1}{x} \,.
\ee
Using this identification and the relations in \eqref{Q0map}, it can be checked that
\begin{align}
\mathcal{W}_{Q_1}(\sigma_\star)\big|_{1\text{-inst}}=\mathcal{W}^{-+}_{1\text{-inst}}
\end{align}
{\it{i.e.}} the 1-instanton contribution of ${\mathcal W}_{Q_1}$ matches the localization result \eqref{LOCJK2} with the $(-+)$ prescription. 
We have checked in several examples that this match also occurs at higher instanton
numbers.

\subsection{The dual theory}
In the previous section we have shown that, with an appropriate map of parameters,
the twisted superpotentials of the quivers $Q_0$ and $Q_1$ match the results 
from localization obtained with two distinct JK prescriptions. Since these differ already at 
the 1-instanton level, as shown in \eqref{JK1JK2difference}, it is clear that the superpotentials 
$\mathcal{W}_{Q_0}$ and $\mathcal{W}_{Q_1}$
{\emph {do not match}}.
By studying a number of low-rank theories at the first few instanton orders, 
we find that the difference between the
superpotentials of the two quivers can be written as
\begin{align}
\label{WQ2Q1powerseries}
{\mathcal W}_{Q_1}(\sigma_\star) - {\mathcal W}_{Q_0}(\sigma_\star) =& ~(-1)^{N+1} \Big
(x+\frac{x^2}{2} + \frac{x^3}{3} + \ldots \Big)\sum_{i\in\mathcal{F}_1}m_i  \cr
& + (-1)^{N+1} \Big(\frac{q_0}{x}+\frac{q_0^2}{2x^2} + \frac{q_0^3}{3x^3} + \ldots \Big)
\sum_{i\in\mathcal{F}_1}m_i  \,.
\end{align}
After using the map \eqref{Q0map} we observe that this is in complete agreement with the resummed result \eqref{DeltaWexact} obtained for the superpotentials calculated from the $(+-)$ and $(-+)$ prescriptions using localization. 
This not only supports our identification between contours and quivers, it also shows us the way to correctly identify dual pairs of quiver theories. 

In fact, given the simple relation \eqref{WQ2Q1powerseries}, 
it is natural to propose that the quiver theory that is actually dual to $Q_0$ and that we denote by $\widetilde{Q}_1$, is the one whose superpotential 
differs from that of $Q_1$ by non-perturbative corrections according to
\begin{align}
\label{WQ2tildeSimple}
{\mathcal W}_{\widetilde{Q}_1}&=-\log x\,\sum_{s\in \mathcal{N}_2}\,\sigma_s-\sum_{i\in \mathcal{F}_1}\sum_{s\in \mathcal{N}_2}\varpi(\sigma_s-m_i)-\sum_{s\in \mathcal{N}_2}\Big\langle\text{Tr}\,\varpi(\Phi-\sigma_s)\Big\rangle\notag\\[2mm]
&\hspace{1.2cm}+\bigg(\log\Big(1-(-1)^{N
}\,x\Big)+\log\Big(1-(-1)^{N} \frac{q_0}{x}\Big)\bigg)\sum_{i\in\mathcal{F}_1}m_i \,.
\end{align}
The first line of the right hand side 
of the equation above is identical to the superpotential of the quiver $Q_1$
given in \eqref{pN-pQ2W}, in which we have used the map \eqref{yvsx2node}.
The second line in (\ref{WQ2tildeSimple}) encodes the non-perturbative corrections to the naive answer. Expanding the logarithm, we see that the power series coincides with the difference calculated in \eqref{WQ2Q1powerseries}.
Furthermore, this is precisely what we derive from first principles using contour deformation arguments in Appendix \ref{contourdeform}. 

The appearance of non-perturbative terms in the superpotential of the dual theory is in part already known in the context of conformal gauge theories in two dimensions. Indeed, as shown in \cite{Benini:2014mia}, 2d Seiberg duality requires not only the inversion of the FI couplings, but also that the twisted superpotential of the dual quiver is modified by the following non-perturbative correction:
\begin{align}
\label{generalizedduality2d}
\delta W=\log\Big(1-(-1)^{N_f}\,x\Big)\, \big(\Tr \widetilde m - \Tr m\big)\,.
\end{align}
where $x$ is the exponentiated FI parameter of the 2d gauge node that is dualized, and $\Tr m$ and $\Tr \widetilde m$ denote respectively the sum of twisted masses for all $N_f$ fundamental and $N_f$ anti-fundamental flavours attached to that node.

The purely 2d part of the non perturbative term 
in \eqref{WQ2tildeSimple} is exactly $\delta W$ in \eqref{generalizedduality2d}, written for the particular case we are considering.
Note that in \eqref{WQ2tildeSimple} it depends only on the anti-fundamental flavours attached to the dualized node in Fig.~\ref{Q1pN-p}, because in $Q_0$ the contribution from fundamental flavours is solely due to the 4d node and this vanishes due to the tracelessness condition of SU$(N)$. Our result shows that, whenever a 2d gauge node connected to a dynamical 4d gauge node is dualized, there is also an extra contribution that arises as a consequence of the non-trivial 4d dynamics. The modified Seiberg rule in \eqref{WQ2tildeSimple} is thus a generalization of the one in \eqref{generalizedduality2d} and represents the main result of this section.

\subsection{Basic rules of duality}

The definition of the dual quiver we have introduced might seem simply a change in nomenclature since the non-perturbative terms we have added 
are constant and do not affect the dynamics or twisted chiral ring equations. However, in a generic quiver
with more nodes, the fundamental or anti-fundamental matter fields of a given 
2d node are realized by other 2d gauge nodes; in this case the role 
of the twisted masses will be played by $\Tr \sigma$ of that gauge node and thus
such terms do affect the dynamics. Indeed, they 
affect the form of the twisted chiral ring equations.

In summary the basic duality rules for the twisted chiral superpotentials of pairs of dual quivers are:
\begin{enumerate}

\item The ranks of the gauge and flavour nodes of the dual quiver are 
completely determined by the operation shown in Figures \ref{PreBasicDuality} and \ref{PostBasicDuality}. 

\item For such a duality move, the exponentiated FI couplings of the pair of dual quivers are related by inversion, as shown in \eqref{yvsx2node}.

\item If the dualized node is only connected to flavour or 
other 2d gauge nodes, the twisted chiral superpotential of the dual quiver is corrected by a non-perturbative piece given in \eqref{generalizedduality2d}. The twisted masses are replaced by the twisted scalars of the vector multiplet in case the flavour is realized by a 2d gauge node. 

\item If the dualized node is connected to the dynamical 4d gauge node, the non-perturbative correction to the twisted superpotential takes the form:
\begin{align}
\label{generalizedduality}
\delta W=\bigg[\log\Big(1-(-1)^{N_f}\,x\Big)+\log\left(1-(-1)^{N_f} \frac{q_0}{x}\right)\bigg]
\big(\Tr \widetilde m - \Tr m\big)\,,
\end{align}
where $N_f$ is the number of (anti-) fundamental flavours attached to the dualized node.
As before, when the flavour symmetry is realized by a 2d gauge node, the twisted masses are replaced by the twisted scalars in the 2d vector multiplet. 
\end{enumerate}
Given these duality rules and the resulting twisted superpotential of the dual quiver theory, we solve the twisted chiral ring equations order by order in the exponentiated FI couplings. Upon evaluating the superpotential on the solutions of the chiral ring equations, we find a perfect match 
with the evaluation of the superpotential on the corresponding massive vacuum of original quiver. 

\section{Seiberg duality for 3-node quivers}
\label{SD3nodes}
\begin{figure}[t]
\begin{center}
{\footnotesize
\begin{tikzpicture}[decoration={
markings,
mark=at position 0.5 with {\draw (-5pt,-5pt) -- (0pt,0pt);
                \draw (-5pt,5pt) -- (0pt,0pt);}}]
  \matrix[row sep=1mm,column sep=3mm] {
     &\node{$\phantom{n_3+}$}; & & \node[](AfNW)[squashedflavor] {$n_2+n_3$};  && \\      
     \node(Ag1)[gauge] {$n_1$};  & & \node(Ag2)[gauge] {$n_1+n_2$}; & &\node(Agf)[gaugedflavor]{$N$};&\\
      & & &[-8mm]\node[xshift=-8mm](AfSW)[squashedflavor] {$n_1+n_2$};  &&\node[xshift=+10mm,yshift=-5mm](AfSE)[squashedflavor] {$n_3+n_1$}; \\
     &\node{$\phantom{n_3+}$}; & & \node[](BfNW)[squashedflavor] {$n_2+n_3$};  && \\      
     \node(Bg1)[gauge] {$n_2$};  & & \node(Bg2)[gauge] {$n_1+n_2$}; & &\node(Bgf)[gaugedflavor]{$N$};&\\
      & & &\node[](BfSW)[squashedflavor] {$n_1+n_2$};  &&\node[xshift=+10mm,yshift=-5mm](BfSE)[squashedflavor] {$n_3+n_1$}; \\
     &\node{$\phantom{n_3+}$}; & & \node[xshift=-5mm](CfNW)[squashedflavor] {$n_2+n_3$};  && \\      
     \node(Cg1)[gauge] {$n_2$};  & & \node(Cg2)[gauge] {$n_2+n_3$}; & &\node(Cgf)[gaugedflavor]{$N$};&\\
      & & &\node[](CfSW)[squashedflavor] {$n_1+n_2$};  &&\node[xshift=+10mm,yshift=-5mm](CfSE)[squashedflavor] {$n_3+n_1$}; \\
     &\node{$\phantom{n_3+}$}; & & \node[xshift=-8mm](DfNW)[squashedflavor] {$n_2+n_3$};  && \\      
     \node(Dg1)[gauge] {$n_3$};  & & \node(Dg2)[gauge] {$n_2+n_3$}; & &\node(Dgf)[gaugedflavor]{$N$};&\\
      & & &\node[](DfSW)[squashedflavor] {$n_1+n_2$};  &&\node[xshift=+10mm,yshift=-5mm](DfSE)[squashedflavor] {$n_3+n_1$}; \\
   };
\graph[edges={postaction={decorate}}]{
(Ag1)--(Ag2)--(Agf);
(AfNW)--[bend right,looseness=.7](Ag2);(AfNW)--[bend left,right anchor=north,looseness=.9](Agf);
(AfSW)--[bend left,out=+12](Ag1);(AfSW)--[bend right,left anchor=east,looseness=.7](Agf);
(AfSE)--(Agf);
(Bg2)--(Bg1);(Bg2)--(Bgf);
(BfNW)--[bend right,looseness=.7](Bg2);(BfNW)--[bend left,right anchor=north,looseness=.9](Bgf);
(Bg1)--[bend right,in=+195](BfSW);(BfSW)--[bend left,out=+15](Bg2);(BfSW)--[bend right,left anchor=east,right anchor=south,looseness=.9](Bgf);
(BfSE)--(Bgf);
(Cg1)--(Cg2);(Cgf)--(Cg2);
(CfNW)--[bend right,out=-20,looseness=.6](Cg1);(Cg2)--[bend left,looseness=.6](CfNW);(CfNW)--[bend left,right anchor=north,looseness=.9](Cgf);
(Cg2)--[bend right](CfSW);(CfSW)--[bend right,left anchor=east,right anchor=south,looseness=.9](Cgf);
(CfSE)--(Cgf);
(Dgf)--(Dg2)--(Dg1);
(Dg1)--[bend left,looseness=.6,right anchor=west,in=-200](DfNW);(DfNW)--[bend left,left anchor=east,right anchor=north,looseness=.7](Dgf);
(Dg2)--[bend right](DfSW);(DfSW)--[bend right,left anchor=east,right anchor=south,looseness=.9](Dgf);
(DfSE)--(Dgf);
};
\begin{scope}[on background layer]
\node(quiverA) [fill=white,fit=(Ag1) (AfSE) (AfNW)] {};
\node[gray] at (quiverA.north west) {\Large ($Q_0$)};
\node(quiverB) [fill=white,fit=(Bg1) (BfSE) (BfNW)] {};
\node[gray] at (quiverB.north west) {\Large ($\widetilde{Q}_1$)};
\node(quiverC) [fill=white,fit=(Cg1) (CfSE) (CfNW)] {};
\node[gray] at (quiverC.north west) {\Large ($\widetilde{Q}_2$)};
\node(quiverD) [fill=white,fit=(Dg1) (DfSE) (DfNW)] {};
\node[gray] at (quiverD.north west) {\Large ($\widetilde{Q}_3$)};
\node(dualin1)[gauge,outer sep=4pt,draw, pattern=crosshatch dots,pattern color=blue!60] at (Ag1){$\phantom{n_1}$};
\node[fill=white] at (Ag1){$\phantom{n_1}$};
\node(dualout1)[gauge,outer sep=4pt,draw, pattern=crosshatch dots,pattern color=red!60] at (Bg1){$\phantom{n_2}$};
\node[fill=white] at (Bg1){$\phantom{n_2}$};
\node(dualin2)[gauge,outer sep=4pt,pattern=crosshatch dots,pattern color=blue!60] at (Bg2){$\phantom{n_1+n_2}$};
\node[fill=white,inner sep=2pt] at (Bg2){$\phantom{n_1+n_2}$};
\node(dualout2)[gauge,outer sep=4pt,draw, pattern=crosshatch dots,pattern color=red!60] at (Cg2){$\phantom{n_2+n_3}$};
\node[fill=white,inner sep=2pt] at (Cg2){$\phantom{n_2+n_3}$};
\node(dualin3)[gauge,outer sep=4pt,align=center,draw, pattern=crosshatch dots,pattern color=blue!60] at (Cg1){$\phantom{n_2}$};
\node[fill=white,align=center] at (Cg1){$\phantom{n_2}$};
\node(dualout3)[gauge,outer sep=4pt,align=center,draw, pattern=crosshatch dots,pattern color=red!60] at (Dg1){$\phantom{n_3}$};
\node[fill=white,align=center] at (Dg1){$\phantom{n_3}$};
\graph[edges={violet,very thick,dashed}]{(dualin1)->(dualout1) (dualin2)->(dualout2) (dualin3)->(dualout3);};
\end{scope}
\end{tikzpicture}
}
\caption{A sequence of dualities relating 2d/4d quivers corresponding to the $[n_1,n_2,n_3]$ defect in SU$(N)$ asymptotically conformal SQCD, starting from the quiver $Q_0$.}
\label{3nodedualitychain}
\end{center}
\end{figure}
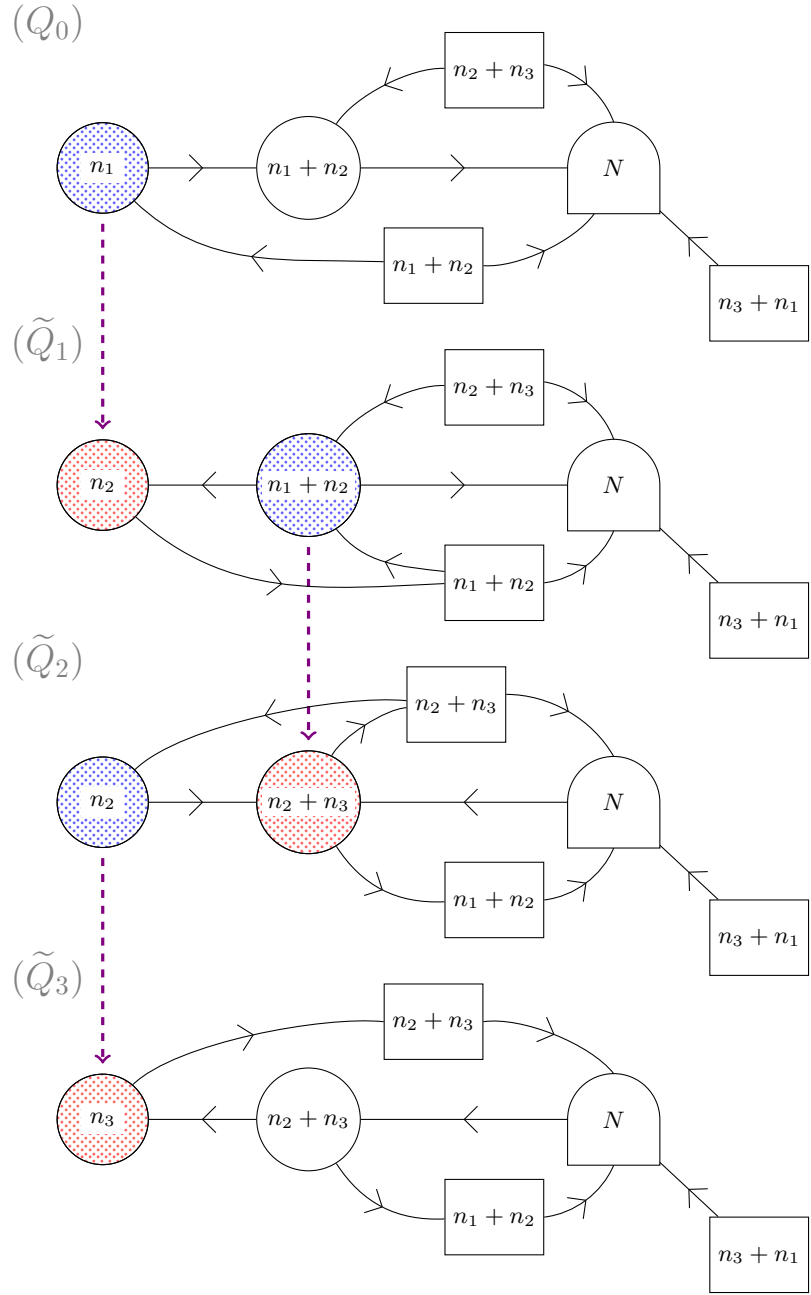

We now
apply the duality rules derived in the previous section to quivers with two gauge nodes and one 
flavour node. 
We begin with the quiver denoted by $Q_0$ and perform the sequence of Seiberg dualities shown in Fig.~\ref{3nodedualitychain}. 

 The ranks and  connectivity of the quivers are determined by the duality rules discussed in Section \ref{GSD}. These are sufficient to determine the classical and one-loop contributions to the twisted chiral superpotential.
With these ingredients alone, starting from $Q_0$ we can obtain
three quivers $Q_{\ell}$ (with $\ell=1,2,3$), whose twisted chiral superpotential $\cW_{Q_\ell}$ is computed, at 1-instanton level, in Appendix \ref{3nodesW}. 
From our previous discussion of Seiberg duality we
know however that each step of the duality chain induces additional non-perturbative corrections for the superpotential. 
We shall therefore use the notation $\widetilde{Q}_{\ell}$, in order to indicate that while their twisted superpotentials share the classical and one-loop parts with those of $Q_{\ell}$, 
they differ by non-perturbative terms. 

The twisted chiral superpotential for the first quiver $Q_0$ is: 
\begin{align}
{\mathcal W}_{Q_0}\big(\{x\}\big)&=\,\log x_1\,\text{Tr}\,\sigma^{(1)}+\log x_2 \,\text{Tr}\,\sigma^{(2)}
-\sum_{s\in\mathcal{N}_1}\sum_{t\in\mathcal{N}_1\cup\mathcal{N}_2}
\varpi(\sigma^{(1)}_s-\sigma^{(2)}_t)\cr
&\hspace{.1cm}-\!\sum_{s\in\mathcal{N}_1}\sum_{i\in\mathcal{F}_1}\varpi(m_i-\sigma^{(1)}_s)
-\!\!\!
\sum_{s\in\mathcal{N}_1\cup\mathcal{N}_2} 
\sum_{i\in\mathcal{F}_2}\varpi(m_i-\sigma^{(2)}_s)-\!\!\!\!
\sum_{s\in\mathcal{N}_1\cup\mathcal{N}_2} \!\!\Big\langle\text{Tr}\,\varpi(\sigma^{(2)}_s-\Phi)
\Big\rangle \,.\cr
\end{align}
We now perform a duality on the U$(n_1)$ gauge node in $Q_0$ to obtain the
quiver $\widetilde Q_1$
whose twisted superpotential is
\begin{align}
{\mathcal W}_{\widetilde Q_1}\big(\{x\}\big)&=-\log x_1\,\text{Tr}\,\sigma^{(1)}+\log(x_1x_2)\,\text{Tr}\,\sigma^{(2)}-\sum_{s\in\mathcal{N}_2}\sum_{i\in\mathcal{F}_1}\varpi(\sigma^{(1)}_s-m_i)\cr
&\hspace{5mm}-\sum_{s\in\mathcal{N}_2}\sum_{t\in\mathcal{N}_1\cup\mathcal{N}_2}\varpi(\sigma^{(2)}_t-\sigma^{(1)}_s)-\sum_{s\in\mathcal{N}_1\cup\mathcal{N}_2}\sum_{i\in\mathcal{F}_1\cup\mathcal{F}_2}\varpi(m_i-\sigma^{(2)}_s)\label{WQtilde1}\\[2mm]
&\hspace{5mm}-\!\!\sum_{s\in\mathcal{N}_1\cup\mathcal{N}_2}\!\!
\Big\langle\text{Tr}\,\varpi(\sigma^{(2)}_s-\Phi)\Big\rangle
+\log\Big(1-(-1)^{n_1+n_2}\,x_1\Big)
\Big(\sum_{i\in\mathcal{F}_1}m_i-\text{Tr}\,\sigma^{(2)}\Big)\,.\notag
\end{align}
%
The last logarithmic term accounts for the non-perturbative corrections due to the
standard duality rule \eqref{generalizedduality2d} in which we have used $N_f=n_1+n_2$
since it is the U$(n_1+n_2)$ gauge node that provides fundamental matter to the 
U$(n_1)$ node that is dualized.

In order to see the effect of the duality more clearly, one can write the above superpotential using the  variables that are natural for 
the quiver $Q_1$, by collecting the $\Tr\sigma^{(2)}$ terms together. 
Comparing with the superpotential ${\mathcal W}_{Q_1}$ given in
\eqref{WQ2y}, we have
\be
{\mathcal W}_{\widetilde Q_1}\big(\{x\}\big)=
{\mathcal W}_{Q_1}\big(\{y\}\big)+\log\Big(1-(-1)^{n_1+n_2}\,x_1\Big)
\, \sum_{i\in\mathcal{F}_1}m_i \,,
\ee
where the FI parameters $(y_1, y_2)$ 
appearing in ${\mathcal W}_{Q_1}$ 
are
\begin{align}
\label{yvsx}
y_1=\frac{1}{x_1}\,,\quad y_2=\frac{x_1x_2}{1-(-1)^{n_1+n_2}\,x_1}\,.
\end{align}
Here we see how the Seiberg duality acts on the FI parameters when more than one gauge node is present
\footnote{As shown in \cite{Benini:2014mia}, it is possible to define cluster variables in terms of which the Seiberg duality action on the FI parameters can be recast as a cluster algebra.}. 

The twisted chiral ring equations obtained from ${\mathcal W}_{Q_0}$ and ${\mathcal W}_{\widetilde{Q}_1}$ can be solved as usual by expanding about a particular classical vacuum that corresponds to the surface operator and performing an order-by-order expansion in the exponentiated FI couplings $x_I$. Upon evaluating the respective superpotentials on the resulting solutions, we find a perfect match up to purely $q_0$-dependent
terms. We have checked this up to 8 (ramified) instantons for several low rank cases and this agreement is a confirmation of the proposal for 2d Seiberg duality at the level of the low energy effective action.

It is important to mention here that the twisted chiral ring equations one would write for $\widetilde{Q}_1$ are different from those that one would write for the quiver $Q_1$ on account of the non-perturbative corrections to the FI parameters of the dual theory. It is only with these corrections that the equality with the low-energy superpotential $\cW_{Q_0}$ holds.

Along the same lines, we now consider the second and third dualities moves in the duality chain in Figure \ref{3nodedualitychain}. In the former, the dualized 2d node is connected to the
4d gauge node, and thus the modified duality rules \eqref{generalizedduality} have to be used.
This duality step leads to the quiver $\widetilde{Q}_2$ and, collecting terms as before, we find 
\begin{eqnarray}
{\mathcal W}_{\widetilde{Q}_2}\big(\{x\}\big) = {\mathcal W}_{Q_2}\big(\{z\}\big) 
&+&\bigg[\log\Big(1-(-1)^{n_1+n_3}\, y_2\Big)
+\log\Big(1-(-1)^{n_1+n_3}\frac{q_0}{ y_2}\Big)\bigg]
\!\sum_{i\in\mathcal{F}_1\cup\mathcal{F}_2}\!\!m_i \nonumber\\[2mm]
&+&\log\Big(1-(-1)^{n_1+n_2} \,x_1\Big)\  \sum_{i\in\mathcal{F}_1}m_i\,.
\end{eqnarray}
The superpotential ${\mathcal W}_{Q_2}$ is defined 
in \eqref{WQ3z} in Appendix~\ref{3nodesW} and is determined purely by the connectivity of the quiver $Q_2$ whose FI parameters we denote 
$(z_1, z_2)$ are expressed in terms of those of the original quiver $Q_0$ via their dependence on
$y_I$ according to
\be
\label{zvsy}
{z}_1 = -\frac{{y}_1{y}_2}{(1-{y}_2)(1-\frac{q_0}{{y}_2})}\,, \quad {z}_2 = \frac{1}{ y_2}\,.
\ee
In \eqref{zvsy} we see 
the appearance of $q_0$
since the dualized node is directly connected to the dynamical 4d node. 

Finally, we perform the third duality move and obtain the quiver denoted by $\widetilde{Q}_3$ in Figure \ref{3nodedualitychain}; its twisted superpotential is:
\begin{align}
{\mathcal W}_{\widetilde{Q}_3}\big(\{x\}\big) = &~{\mathcal W}_{Q_3}\big(\{w\}\big) 
+ \log\Big(1-(-1)^{n_2+n_3}\,{z}_1\Big)\sum_{i\in {\mathcal F}_2} m_i\cr
&~~~+\bigg[\log\Big(1-(-1)^{n_1+n_3}\, y_2\Big)+
\log\Big(1-(-1)^{n_1+n_3}\,\frac{q_0}{ y_2}\Big)\bigg]
\sum_{i\in\mathcal{F}_1\cup\mathcal{F}_2} \!m_i \cr
&~~~+\log\Big(1-(-1)^{n_1+n_2}\, x_1\Big)\sum_{i\in\mathcal{F}_1} m_i \,.
\end{align}
The superpotential ${\mathcal W}_{Q_3}$ is defined in \eqref{WQ4w} in Appendix~\ref{3nodesW}
and its FI parameters $({w}_1, {w}_2)$ are
\be
{w}_1 = \frac{1}{{z}_1}\, \quad {w}_2 = \frac{{z}_1{z}_2}{1-(-1)^{n_2+n_3}\,{z}_1}\,.
\ee
By successively composing the relations \eqref{zvsy} and \eqref{yvsx}, one can express these FI couplings in terms of those of the original
quiver $Q_0$. 

Once the twisted chiral superpotentials of the dual quivers are obtained, we can solve the corresponding chiral ring relations as usual and evaluate the superpotentials on these solutions. Our calculations confirm the equality of these quantities and show that, up to purely $q_0$-dependent terms, the three quivers $\widetilde{Q}_{1,2,3}$ derived from $Q_0$, lead to the same low energy effective action on the Coulomb branch.

\subsection{Rows of dual quivers}

So far, we have worked out a single duality chain starting with the quiver $Q_0$ and shown that the twisted superpotentials evaluated on the solutions to the twisted chiral ring equations match for all these four 3-node quivers:
\be
Q_0 \longleftrightarrow \widetilde{Q}_1 \longleftrightarrow \widetilde{Q}_2 \longleftrightarrow \widetilde{Q}_3 \,.
\ee
The arrows are double headed since dualities can be performed in either direction. The new result has been the second duality move, which is a generalized duality and involves a change in the superpotential as shown in \eqref{generalizedduality}. 

These results can be easily generalized to the generic case in which the gauge nodes form a linear quiver. Our earlier work has shown that, for such an $M$-node case, there are $2^{M-1}$ possible Seiberg-dual quivers \cite{Ashok:2018zxp}. Each such quiver is labelled by a vector $(s_1, s_2, \ldots, s_{M-1})$ whose entries take values $0$ or $1$. For instance, for the 3-node cases studied in this work, we find the following set of quivers that map to distinct JK vectors on the localization side, that are completely determined by the permutation $\vec{s}$ (see \cite{Ashok:2018zxp} for details):
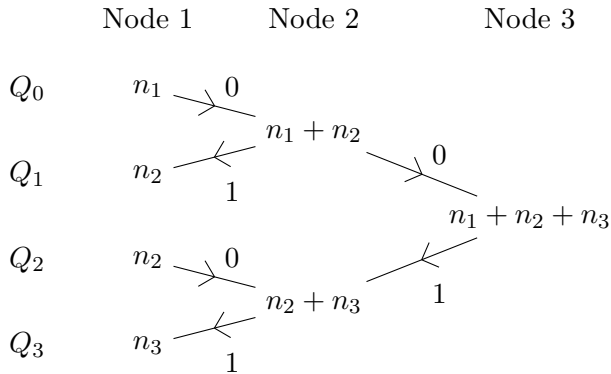
\begin{figure}[H]
\begin{centering}
\begin{tikzpicture}[decoration={
markings,
mark=at position 0.5 with {\draw (-5pt,-5pt) -- (0pt,0pt);
                \draw (-5pt,5pt) -- (0pt,0pt);}}]
  \matrix(table)[column sep=5mm] {
      && \node{Node 1};&[+.5em]   \node{Node 2}; &[+1em] \node{Node 3};
      \\ [+1em]     
      &\node{$Q_0$};& \node(q1n1){$n_1$};&  &  \\
      &&  &\node(q12n2){$n_1+n_2$};  & & \\
      &\node{$Q_1$};& \node(q2n1){$n_2$};&  &  \\
      && &  &  \node(qtopn3){$n_1+n_2+n_3$}; \\
      &\node{$Q_2$};& \node(q3n1){$n_2$};&  &  \\
      && & \node(q32bn2){$n_2+n_3$};  &  \\
      &\node{$Q_3$};& \node(q2bn1){$n_3$};&  &  \\
             };
      
  \graph[edge label=$0$,edges={postaction={decorate}}]{
  (q1n1)--(q12n2); (q3n1)--(q32bn2); 
  (q12n2)--(qtopn3);
  };

 \graph[edge label=$1\phantom{\Big)}$,edges={postaction={decorate}}]{
  (q12n2)--(q2n1); (q32bn2)--(q2bn1); 
  (qtopn3)--(q32bn2); 
  };
\end{tikzpicture}
\caption{The linear 3-node quivers that are Seiberg-dual to the oriented quiver $Q_0$. Only the gauge nodes are shown, the flavour nodes can be assigned unambiguously such that each 2d gauge node is conformal. 
The $s_I$ that label the quiver are drawn on the arrows linking the gauge nodes.}
\label{4nodelinearlist}
\end{centering}
\end{figure}
For the asymptotically conformal gauge theories, as we have seen, for each duality move, one has to add non-perturbative corrections in order to obtain the correct twisted superpotential of the dual quiver. So, given a dual quiver specified by a permutation $\vec{s}$, there are two steps to be carried out: first, one needs to find out the sequence of Seiberg-duality moves needed to connect the quiver $Q_0$ to any one of the quivers in the list. Secondly, one has to add appropriate non-perturbative corrections after each duality move. 

The way Seiberg duality moves are encoded in terms of the permutation basis
can be described by realizing that there are only $M-1$ basic duality moves, that correspond to dualizing one of the $M-1$ 2d gauge nodes. 
Given that the $M-1$ arrows of the quivers are also denoted by the same vector $\vec{s}$, and knowing the action of duality, which exchanges fundamental with anti-fundamental matter on the dualized node, it then follows that the basis of duality moves can be represented by the following actions on the vector $\vec{s}$:
\begin{align}
{\cal D}_1 &: (***\ldots 0) \rightarrow (***\ldots 1) \cr
{\cal D}_2 &: (***\ldots 10) \rightarrow (***\ldots 01)\cr
{\cal D}_3 &: (**\ldots 10*) \rightarrow (**\ldots 01*) \quad\text{and so on}\,.
\end{align}
In this way, it is easy to find out how any quiver labelled by $\vec{s}$ can be connected to $Q_0$ by a sequence of duality moves. Once this is done, one can add the appropriate non-perturbative corrections to the twisted superpotential after each duality
using the rules explained in Section \ref{GSD} and obtain a row of dual theories, just as before:
\be
Q_0\stackrel{ {\cal D}_1}{\longleftrightarrow} \widetilde{Q}_1\stackrel{{\cal D}_2}{ \longleftrightarrow} \widetilde{Q}_2 \longleftrightarrow  \cdots \,.
\ee
This solves the problem of finding dual quivers related to $Q_0$ for the generic linear quiver. 

We conclude with the following observation: given the localization integrand, one could choose any JK prescription to evaluate the partition function. On the 2d/4d quiver side, this corresponds to choosing a particular quiver $Q_k$; one could then perform a set of Seiberg dualities:  
\begin{align}
\label{dualitymatrix}
&\widehat{Q}_0 \stackrel{ {\cal D}_1}{\longleftrightarrow} \widehat{Q}_1 \stackrel{ {\cal D}_2}{\longleftrightarrow} \widehat{Q}_2 \cdots 
 \longleftrightarrow Q_k \longleftrightarrow \widehat{Q}_{k+1} \longleftrightarrow\cdots 
\end{align}
In this duality chain, the quiver $Q_k$ has a Lagrangian that one would write down purely from the quiver itself. All the others $\widehat{Q}_{\ell}$ are related to it by Seiberg-dualtiy and their superpotentials would differ from those one would write for the quiver $Q_{\ell}$ by non-perturbative pieces determined by the sequence of dualities involved. The low energy superpotentials for each quiver in the chain are identical to that obtained for $Q_k$ (up to purely $q_0$-dependent terms).  One can therefore write down $2^{M-1}$ such duality chains starting with any of the quivers corresponding to a given JK prescription. The results match along the rows of dual quiver: these are interpreted as 
the result of deforming the integration contour from one set of poles to another, keeping into account the residues at infinity.
%
\vspace{1cm}
\section*{Acknowledgements}

We thank Marco Bill\`o and Alberto Lerda for collaboration at an initial stage of the work. We thank E. Dell'Aquila, J. Distler, S. Gukov, V. Gupta, D. Jatkar, and M. Raman for useful discussions, and Alberto Lerda for a very careful reading of the manuscript and for helpful comments and suggestions. 

The work of M.F. and R.R.J. is partially supported by the MIUR PRIN Contract 
2015MP2CX4 ``Non-perturbative Aspects Of Gauge Theories And Strings''. 
\begin{appendix}
\section{Contour deformations}
\label{contourdeform}
In this section, we derive the 2d/4d Seiberg duality rule in SQCD by lifting the theory to three dimensions and studying the partition function of the surface operator of type $[n_1, n_2]$ with support $\mathbb{R}^2\times S^1_\beta$ in $\mathbb{R}^4\times S^1_{\beta}$. As we shall see, the extra circle direction allows us to relate the partition functions for the $(+-)$ and $(-+)$ contours up to all orders in the instanton expansion. We follow the basic ideas in \cite{Hwang:2017kmk} though we will keep the 4d instanton weight $q_0\ne 0$. In the end we will take the four dimensional limit $\beta\rightarrow 0$ and set the $\Omega$-deformation parameters $\epsilon_i$ to zero in order to read off how the 2d twisted superpotentials obtained using the two prescriptions are related.

The partition function for the $(+-)$ contour in the 2-node case is given by 
\be
Z^{+-} = \sum_{d_1, d_2}  \frac{(-q_1)^{d_1}}{d_1!} \frac{(-q_2)^{d_2}}{d_2!} \int_+ \prod_{\sigma=1}^{d_1} \frac{d\chi_{1,\sigma}}{2\pi\ii} \int_- \prod_{\rho=1}^{d_2} \frac{d\chi_{2,\rho}}{2\pi\ii}\, z_{\{d_I\}} \,,
\ee
where the integrand takes the following form:
\begin{align}
z_{\{d_I\}} & = \,\prod_{I=1}^2\prod_{\sigma,\tau=1}^{d_I}\,
\frac{\sinh\frac{\beta}{2}\left(\chi_{I,\sigma} - \chi_{I,\tau} + \delta_{\sigma,\tau}\right)
}{\sinh\frac{\beta}{2}\left(\chi_{I,\sigma} - \chi_{I,\tau} + \epsilon_1\right)}\cr
&\times \prod_{\sigma=1}^{d_1}\prod_{\rho=1}^{d_2}\,
\frac{\sinh\frac{\beta}{2}\left(\chi_{1,\sigma} - \chi_{2,\rho} + \epsilon_1 + \hat\epsilon_2\right)}
{\sinh\frac{\beta}{2}\left(\chi_{1,\sigma} - \chi_{2,\rho} + \hat\epsilon_2\right)}
\frac{\sinh\frac{\beta}{2}\left(\chi_{2,\rho} - \chi_{1,\sigma} + \epsilon_1 + \hat\epsilon_2\right)}
{\sinh\frac{\beta}{2}\left(\chi_{2,\rho} - \chi_{1,\sigma} + \hat\epsilon_2\right)}
\label{zexplicit3d5d}
\\
& ~~\times
\prod_{I=1}^2 \prod_{\sigma=1}^{d_I} \frac{\prod_{i\in {\mathcal F}_I} \sinh\frac{\beta}{2}(\chi_{I,\sigma} - m_{i})}
{\prod_{s\in{\mathcal N}_I}\sinh\frac{\beta}{2}\left(a_{s}-\chi_{I,\sigma} + \frac 12 (\epsilon_1 + \hat\epsilon_2)\right)
\prod_{t\in{\mathcal N}_{I+1}}
\sinh\frac{\beta}{2}\left(\chi_{I,\sigma} - a_{t} + \frac 12 (\epsilon_1 + \hat\epsilon_2)\right)}~.
\nonumber
\end{align}
This is obtained from the integrand in \eqref{zexplicit5d} by lifting rational functions to trigonometric functions. Since we are eventually interested only in the strict 4d limit, we have not turned on either 3d or 5d Chern-Simons levels. Given this starting point, our goal is to deform the contour to obtain $Z_{-+}$, knowing that as emphasized in \cite{Hwang:2017kmk}, due to a non-trivial residue at infinity, one should obtain a wall-crossing type pre-factor. 

Let us review how this works for the case in which $q_2=0$. The second line in \eqref{zexplicit3d5d} is not present in such a case and only terms with $I=1$ survive.  
The integral receives contributions from multiple residues at the singularities of the integrand. When we reverse the contour, among the possible singularities to consider there are both the ones at finite points and the ones at infinity. In the particular case of our integrand, there are residues at both asymptotes $\pm \infty$. 
Out of the $d_1$ integration variables, let us assume that $p_1$ of them are evaluated at their poles in the asymptotes. We will eventually sum over all values of $p_1$ from 0 up to $d_1$. The integrand breaks up naturally into three sets of terms: the first involves just the $p_1$ variables that approach infinity; a second, which involves only the complementary set and a last piece, which involves both; after taking the limit in which the $\chi_{1,\sigma}$ are taken in this last piece, we find the following result for the residue:
\begin{align}
\text{Res}_{(+)}z_{d_1} &= \sum_{p=0}^{d_1}\text{Res}_{\cap_\sigma \chi_{1,\sigma}\in \text{Asymp}\pm}\left[ \prod_{\sigma,\tau=1}^{p_1}\,
\frac{\sinh\frac{\beta}{2}\left(\chi_{1,\sigma} - \chi_{1,\tau} + \delta_{\sigma,\tau}\right)
}{\sinh\frac{\beta}{2}\left(\chi_{1,\sigma} - \chi_{1,\tau} + \epsilon_1\right)}\right] \cr
&\hspace{2cm}\times (-1)^{p_1 n_1} e^{-\frac{\beta}{2}p_1\left(\sum_{j\in\mathcal{F}_1} m_j-\sum_{u\in\mathcal{N}_1\cup\mathcal{N}_2} a_u\right) }\,  \text{Res}_{(-)}z_{d_1-p_1}\,.
\end{align}
The sum over $a_u$ gives zero due to the tracelessness condition. The way to deal with the residue coming from the asymptotic region is identical to what is calculated in \cite{Hwang:2017kmk} and we refer the reader to that reference for the details. 
The final result for the case when there are equal numbers of fundamental and anti-fundamental 2d flavours is given as follows (see equation (4.44) of \cite{Hwang:2017kmk}):
\be
Z^{+} = Z^{-} \times \text{PE} \left[ 
\frac
{
2(-1)^{n_1}q_1\left( 
e^{-\frac{\beta}{2} \sum_{i\in\mathcal{F}_1} m_i} - e^{+\frac{\beta}{2} \sum_{i\in\mathcal{F}_1} m_i} 
\right)
}
{(1-e^{2\beta\epsilon_1})}
\right]
\ee
Here we have written the result in terms of the plethystic exponential. For a function $f(t)$ given by  a series expansion:
\be
f(q_1) =\sum_{n=0}^{\infty} f_n q_1^n \Longrightarrow \text{PE}\left[ f(q_1)\right] = \frac{1}{\prod_{n=1}^{\infty} (1-q_1^n)^{f_n}} \,.
\ee
For our case, the function whose plethystic exponential is taken is a linear one and we consider a series expansion in $(-1)^{n_1+1}q_1$.
In order to understand what this means for the superpotential that governs 4d effective action, we take the $\beta\rightarrow 0$ limit and find
%
\be
Z^{+} =Z^{-}\times \text{PE}\left[ (-1)^{n_1} q_1 \frac{1}{\epsilon_1} \sum_{i\in\mathcal{F}_1} m_i \right] 
\ee
The $\epsilon_1\rightarrow 0$ limit then allows one to extract the low energy twisted chiral superpotential from the partition function via\footnote{Since $q_0=0$, the prepotential is zero.}
\be
\lim_{\epsilon_1\rightarrow 0} Z^{\pm} = e^{\frac{\mathcal W^{\pm}}{\epsilon_1}} \,.
\ee
Putting all this together, we find that
\be
\label{formulaAppendixA}
{\mathcal W}^{+} - {\mathcal W}^{-} = \log(1+(-1)^{n_1}q_1) \sum_{i\in\mathcal{F}_1} m_i \,.
\ee
Using the relation  \eqref{Q0map} between the vortex counting parameter $q_1$ and the exponentiated FI parameter of the quiver $Q_0$,
%
%
we find that \eqref{formulaAppendixA} is the same result proposed in \cite{Benini:2014mia} for the twisted chiral superpotentials of quivers related by Seiberg duality using the $S^2$ partition function. Here we have shown that this can be derived from a simple contour deformation argument. 

We now turn to generalize this to the case in which $q_0\ne 0$, starting from the integrand in \eqref{zexplicit3d5d}. We begin with the $(+-)$ contour and deform it to the $(-+)$ contour; in the deformation process, one picks up contributions from the asymptotes $\chi_{I,\sigma} \rightarrow \pm \infty$. 
Let us consider the term in which, out of the $(d_1, d_2)$ integration variables, we let $(p_1, p_2)$ of them to approach infinity. As before, the integrand breaks up into three sets of terms: the first involves only those $\chi_{I,\sigma}$ that take asymptotic values; another set that involves the complementary $\chi_{I,\sigma}$ that take finite values and lastly, those that take values in both sets. After taking the asymptotic limit in this last piece and summing over all the possible values for $p_1$ and $p_2$, we find the following result for the residue:
\begin{align}
\text{Res}_{(+-)}z_{d_1,d_2} &= \sum_{p_1=0}^{d_1}\,\sum_{p_2=0}^{d_2}\text{Res}_{\cap_\sigma \chi_{1,\sigma}\in \text{Asymp}\pm}\left[\prod_{I=1}^2 \prod_{\sigma,\tau=1}^{p_1}\,
\frac{\sinh\frac{\beta}{2}\left(\chi_{I,\sigma} - \chi_{I,\tau} + \delta_{\sigma,\tau}\right)
}{\sinh\frac{\beta}{2}\left(\chi_{I,\sigma} - \chi_{I,\tau} + \epsilon_1\right)}\right.\cr
&\left.\times \prod_{\sigma=1}^{p_1}\prod_{\rho=1}^{p_2}\,
\frac{\sinh\frac{\beta}{2}\left(\chi_{1,\sigma} - \chi_{2,\rho} + \epsilon_1 + \hat\epsilon_2\right)}
{\sinh\frac{\beta}{2}\left(\chi_{1,\sigma} - \chi_{2,\rho} + \hat\epsilon_2\right)}
\frac{\sinh\frac{\beta}{2}\left(\chi_{2,\rho} - \chi_{1,\sigma} + \epsilon_1 + \hat\epsilon_2\right)}
{\sinh\frac{\beta}{2}\left(\chi_{2,\rho} - \chi_{1,\sigma} + \hat\epsilon_2\right)}
\right] 
\cr
&\times (-1)^{p_1 n_1} e^{-\frac{\beta}{2}p_1\sum_{i\in\mathcal{F}_1} m_i}\, 
 (-1)^{p_2 n_2} e^{-\frac{\beta}{2}p_2 \sum_{i\in\mathcal{F}_2} m_i }\,\text{Res}_{(-+)}z_{d_1-p_1,d_2-p_2}\,. \cr
\end{align}
In the asymptotic residue, there is now  a mixed term between the $\chi_{1,\sigma}$ and $\chi_{2,\rho}$; however, the key observation is that we are only interested in how the twisted chiral superpotential changes across the contour deformation and not the whole partition function, which is obtained by setting $\epsilon_2\rightarrow 0$. In this limit, the mixed term is an even function of $\chi_{1,\sigma}-\chi_{2,\rho}$ and does not lead to any new pole that might contribute to the twisted chiral superpotential. As a result, the residue calculation factorizes into a contribution from the $\chi_{1, \sigma}$ integrals and that from the $\chi_{2,\rho}$ integrals; the calculation for each set is identical to that done for the purely 2d case and we  obtain in the 4d limit,
\be
Z^{+-} = Z^{-+}\times \text{PE}\left[ (-1)^{n_1} q_1 \frac{1}{\epsilon_1} \sum_{i\in\mathcal{F}_1} m_i \right] \times \text{PE}\left[ (-1)^{n_2} q_2 \frac{1}{\epsilon_1} \sum_{i\in\mathcal{F}_2} m_i \right] 
\ee
By using the formula for the plethystic exponential, the tracelessness of the flavour group SU$(2N)$, and the form of the instanton partition function in the limit $\epsilon_1\rightarrow 0$ we finally obtain
\be
{\mathcal W}^{+-} - {\mathcal W}^{-+} = \big(\log(1+(-1)^{n_1}q_1) +\log(1+(-1)^{n_2}q_2) \big) \sum_{i\in\mathcal{F}_1} m_i\,.
\ee
Using the map between the $q_I$ and the exponentiated FI parameters and the 4d couplings we derived in \eqref{Q0map}, and by identifying the $(+-)$ and $(-+)$ contours with the corresponding quivers, we derive the following rule for how the twisted superpotential transforms under the action of Seiberg duality:
\be
\delta W=\left[\log(1-(-1)^{N_f}\,x)+\log\left(1-(-1)^{N_f} \frac{q_0}{x}\right)\right] \sum_{i\in\mathcal{F}_1} m_i  \,,
\ee
where $N_f$ denotes the number of (anti-) fundamental flavours attached to the 2d gauge node. 

\section{4d corrections to the 2d Lagrangian}
\label{resolvent}

In this section we show how to evaluate the 4d instanton corrections to the 2d 
twisted chiral superpotential
due to the presence of the chiral correlator $\langle \Tr \varpi(\sigma-\Phi)\rangle$. 
We write this function as follows: 
\be
\label{integraloflog}
\left\langle\, \Tr\, \varpi(z-\Phi)\, \right\rangle = \int
^z\, dz'\, \left\langle\, \Tr\, \log \frac{(z'-\Phi)}{\mu}\, \right \rangle\,.
\ee
We observe that the 4d observable on the R.H.S. is itself the integral of the generating function of the chiral correlators in the 4d gauge theory, and is referred to as the resolvent of the 4d theory. So we begin with a brief review of known results regarding the resolvent of the ${\mathcal N}=2$ supersymmetric SQCD gauge theory (we follow the discussion in \cite{Billo:2012st}). We then show how the quantum gauge polynomial can be written in terms of the chiral correlators of the gauge theory and finally we show how the 2d Lagrangian is affected by the coupling to the four dimnensional theory. 

\subsection{Resolvents and chiral correlators in 4d asymptotically conformal SQCD}

The Seiberg-Witten curve of the asymptotically conformal SU$(N)$ gauge theory with $N_f=2N$ fundamental flavours is given by
\be
Y^2 = \widehat P(z)^2 - g^2 B(z)\,,
\ee
where the characteristic 
gauge polynomial is given by
\be
\label{quantumgp}
\widehat P(z) = z^N + u_2 z^{N-2} + \ldots + (-1)^Nu_N \,,
\ee
and the flavour polynomial is given by
\be
B(z) = \prod_{i=1}^{2N}(z-m_i) \,.
\ee
The constant $g^2$ is related to the Nekrasov counting parameter $q_0$ by 
\be
g^2 = \frac{4q_0}{(1+q_0)^2} \,.
\ee
The Seiberg-Witten differential is given by
\be
\lambda_{SW} = z\, \frac{d\Psi(z)}{dz} dz\,,
\ee
where the function $\Psi(z)$ is 
\be
\Psi(z) = \log\left(\frac{\widehat P(z) + Y}{\mu^N}\right)\,.
\ee
%
The chiral correlators of the gauge theory $\left\langle\Tr \Phi^{\ell}\right\rangle$
%
can be obtained by expanding (for large $z$) the resolvent:
\be
\left\langle \Tr \frac{1}{z-\Phi}\right\rangle = \frac{d\Psi(z)}{dz} \,.
\ee
Integrating with respect to $z$, we find that the integral of the resolvent has a simple form in terms of the function $\Psi(z)$:
\begin{align}
\label{finalresolvent}
\left\langle \Tr \log \frac{z-\Phi}{\mu} \right\rangle = \log\left((1+q_0)\frac{\widehat P+Y}{2\mu^N}\right) \,.
\end{align}
The constant log piece added on the R.H.S ensures that the large-$z$ expansion of both sides match.

\subsection{Chiral correlators vs. quantum gauge polynomial}

Given the gauge polynomial in \eqref{quantumgp} and 
using
equation \eqref{finalresolvent}, it is possible to write the 
coefficients that appear in the gauge polynomial in terms of the chiral correlators of the quantum gauge theory,
which can be calculated from first principles using localization methods \cite{Bruzzo:2002xf,Losev:2003py,Flume:2004rp,Billo:2012st,Ashok:2016ewb}. Unlike the case of pure gauge theory, in the asymptotically conformal case, this relation is subtle due to the presence of the dimensionless coupling $q_0$ that appears non-trivially in the resolvent. To extract this relation,
it is convenient to 
use an equivalent expression for the resolvent \cite{Billo:2012st}: 
\be\label{finalresolventv2}
\left\langle \Tr \log \frac{z-\Phi}{\mu} \right\rangle = \frac{1}{2} \log\frac{\widehat P(z)+Y}{\widehat P(z)-Y}  + \frac{1}{2} \log \frac{B(z)}{\mu^{2N}} +\frac{1}{2}\log q_0 \,.
\ee
%
Expanding
the R.H.S of \eqref{finalresolventv2} for large $z$ and 
equating
the coefficients of $z^{-k}$ on both sides of the equation
allows us to 
express 
the $u_k$ purely in terms of the $\langle \Tr\Phi^k\rangle$.
In order to write down compact expressions, we express the flavour polynomial also in terms of the symmetric polynomials of the masses $S_k$:
\be
B(z) = z^{2N} + \sum_{j=2}^{2N} (-1)^j \, S_j\, z^{2N-j} \,.
\ee
For the lowest orders, following this procedure, we find:
\begin{align}
\label{ukTrPhik}
u_2 &=-\frac{1}{2}\left(\frac{1-q_0}{1+q_0} \right) \langle \Tr \Phi^2\rangle+\frac{q_0}{1+q_0} S_2\cr 
u_3&= +\frac{1}{3}\left(\frac{1-q_0}{1+q_0} \right) \langle \Tr \Phi^3\rangle+\frac{q_0}{1+q_0} S_3 \cr
u_4&= -\frac{1}{4}\left(\frac{1-q_0}{1+q_0} \right) \langle \Tr \Phi^4 \rangle+\frac{1}{2}\langle \Tr \Phi^2 \rangle\left(\frac{1}{4}\langle \Tr \Phi^2 \rangle+\frac{q_0\, S_2}{1+q_0} \right)+\frac{q_0}{1+q_0} S_4\cr
u_5&= +\frac{1}{5}\left(\frac{1-q_0}{1+q_0} \right) \langle \Tr \Phi^5 \rangle-\frac{1}{3}\langle \Tr \Phi^3 \rangle\left(\frac{1}{2}\langle \Tr \Phi^2 \rangle+\frac{q_0\, S_2}{1+q_0} \right) +\frac{q_0\, S_3}{2(1+q_0)}\langle \Tr \Phi^2 \rangle+\frac{q_0}{1+q_0} S_5\cr
u_6&=  -\frac{1}{6}\left(\frac{1-q_0}{1+q_0} \right) \langle \Tr \Phi^6 \rangle+\frac{1}{4}\langle \Tr \Phi^4 \rangle\left(\frac{1}{2}\langle \Tr \Phi^2 \rangle + \frac{q_0 S_2}{1+q_0} \right)+\frac{1}{3}\langle \Tr \Phi^3 \rangle\left(\frac{1}{6}\langle \Tr \Phi^3 \rangle - \frac{q_0 S_3}{1+q_0} \right)  \cr
&\hspace{.3cm} -\frac{1}{2}\left(\frac{1-q_0}{1+q_0} \right)\langle \Tr \Phi^2 \rangle \left(\frac{1}{24}\left( \langle\Tr \Phi^2 \rangle\right)^2 -\frac{q_0 S_2}{4(1-q_0)}\langle \Tr \Phi^2 \rangle - \frac{q_0}{(1-q_0)} S_4 \right) + \frac{q_0}{1+q_0} S_6 \,. \cr
\end{align}

\subsection{Weak coupling expansions and 4d corrections to the 2d superpotential}

In this section we expand the resolvent in 
\eqref{finalresolventv2} 
as an expansion in small $q_0$. As we have seen, the coefficients $u_k$ in the gauge polynomial $P(z)$ have a $q_0$-expansion; let us formally expand the gauge polynomial as follows:
\be
\widehat P(z) = P(z) + \sum_{n=1}^{\infty} p_{n}(z) q_0^n \,,
\ee
where $P(z)$ is the classical gauge polynomial defined in \eqref{CGP},
%
%
and the $p_n(z)$'s can be calculated using \eqref{ukTrPhik}. Then we find that the resolvent has the following instanton expansion (we suppress the $z$-dependence of the polynomials in order to have compact expressions): 
\begin{align}\label{finalresolventexp}
\left\langle \Tr \log \frac{z-\Phi}{\mu} \right\rangle =  
\log \frac{P}{\mu} &+q_0\left(1 +\frac{p_1}{P} -\frac{B}{P^2} \right)\cr
&+q_0^2 \left(-\frac{3 B^2}{2 P^4}+\frac{2 B \left(P+p_1\right)}{P^3}-\frac{P^2-2 p_2 P+p_1^2}{2 P^2}\right)+\ldots \,.\cr
\end{align}
Substituting this into \eqref{integraloflog} and performing the integral, we obtain the one and two instanton corrections to the twisted superpotential of the 2d quiver due to the 4d theory.

\section{Chiral ring equations and superpotentials at the 1-instanton level}
\label{3nodesW}
In this section we study the four quivers shown in Figure~\ref{3nodedualitychain} in turn, write down the twisted chiral ring equations and calculate the 1-instanton result for the low energy superpotential, which is the evaluation of the twisted chiral superpotential in a particular vacuum. Given these results one can check explicitly that the low energy superpotential for the distinct quivers are different already at the 1-instanton level. 

\subsection*{Quiver $Q_0$}
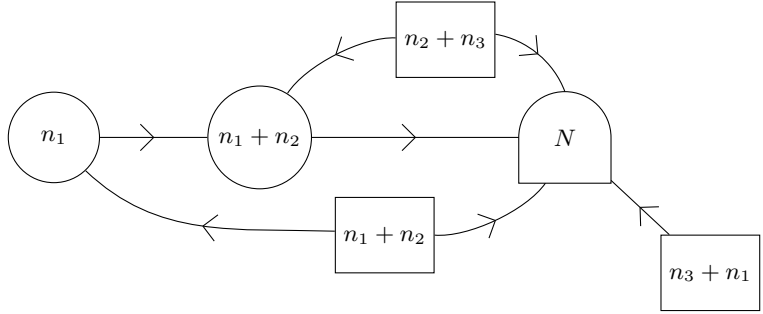
\begin{figure}[H]
\begin{center}
{\footnotesize
\begin{tikzpicture}[decoration={
markings,
mark=at position 0.5 with {\draw (-5pt,-5pt) -- (0pt,0pt);
                \draw (-5pt,5pt) -- (0pt,0pt);}}]
  \matrix[row sep=1mm,column sep=3mm] {
     &\node{$\phantom{n_3+}$}; & & \node[](AfNW)[squashedflavor] {$n_2+n_3$};  && \\      
     \node(Ag1)[gauge] {$n_1$};  & & \node(Ag2)[gauge] {$n_1+n_2$}; & &\node(Agf)[gaugedflavor]{$N$};&\\
      & & &[-8mm]\node[xshift=-8mm](AfSW)[squashedflavor] {$n_1+n_2$};  &&\node[xshift=+10mm,yshift=-5mm](AfSE)[squashedflavor] {$n_3+n_1$}; \\
  };
\graph[edges={postaction={decorate}}]{
(Ag1)--(Ag2)--(Agf);
(AfNW)--[bend right,looseness=.7](Ag2);(AfNW)--[bend left,right anchor=north,looseness=.9](Agf);
(AfSW)--[bend left,out=+12](Ag1);(AfSW)--[bend right,left anchor=east,looseness=.7](Agf);
(AfSE)--(Agf);
};
\end{tikzpicture}
}
\label{Q1App}
\end{center}
\end{figure}
The twisted superpotential is
\begin{align}
\mathcal{W}_{Q_0}(x)=&\log x_1\sum_{s\in {\mathcal N_1}}\sigma^{(1)}_s+\log x_2\sum_{s\in {\mathcal N}_1\cup {\mathcal N}_2}\sigma^{(2)}_s\cr
&-\sum_{t\in {\mathcal N}_1\cup {\mathcal N}_2}\sum_{s\in {\mathcal N_1}}\varpi(\sigma^{(1)}_s-\sigma^{(2)}_t)
-\sum_{i\in {\mathcal F_1}}\sum_{s\in {\mathcal N_1}}\varpi(m_i-\sigma^{(1)}_s)\cr
&-\sum_{i\in {\mathcal F_2}}\sum_{s\in {\mathcal N}_1\cup {\mathcal N}_2}\varpi(m_i-\sigma^{(2)}_s)
-\sum_{s\in {\mathcal N}_1\cup {\mathcal N}_2}\langle\text{Tr}\,\varpi(\sigma^{(2)}_s-\Phi)\rangle\,.
\end{align}
The chiral ring equations are:
\begin{align}
\label{3nodeQ1TCR}
G_2 (\sigma_s^{(1)}) &=\,(-1)^{n_1+n_2}\,\,x_1 B_1 (\sigma_s^{(1)})\quad\text{for}\quad s\in {\mathcal N_1}\,,\cr
(1+q_0)\widehat P(\sigma_s^{(2)})& =\,(-1)^{N}\, \left(x_2\, G_1(\sigma_s^{(2)})B_2(\sigma_s^{(2)})  + \frac{q_0}{x_2} \frac{B_1(\sigma_s^{(2)})B_3(\sigma_s^{(2)})}{G_1(\sigma_s^{(2)})} \right)\,\text{for}\,\, s\in {\mathcal N}_1\cup {\mathcal N}_2\,,\cr
\end{align}
where $G_1(z)$ and $G_2(z)$ are the 2d gauge polynomials for the quiver. We solve the equations about the following vacuum:
\begin{align}
\sigma_s^{(1)} &= a_{s} \quad\text{for}\quad s\in {\mathcal N_1}\,,\cr
\sigma_s^{(2)} &=  a_{s} \quad\text{for}\quad s\in {\mathcal N}_1\cup {\mathcal N}_2\,.
\end{align}
Then at 1-instanton the twisted superpotential evaluated on the solution is
\begin{align}
\mathcal{W}\big|_{\sigma_\star}&=(-1)^{n_1+n_2}x_1\sum_{s\in\mathcal{N}_1}\frac{B_1(a_s)}{P_1'(a_s)P_2(a_s)}
+(-1)^{N+1}x_2\sum_{s\in\mathcal{N}_2}\frac{B_2(a_s)}{P_2'(a_s)P_3(a_s)}\cr
&\hspace{.5cm}+(-1)^{n_3}\frac{q_0}{x_1x_2}\sum_{s\in\mathcal{N}_1}\frac{B_3(a_s)}{P_3(a_s)P_1'(a_s)}\,.
\end{align}
This matches the 1-instanton results from localization at 1-instanton using the contour $(++-)$ if we use the map:
\begin{align}
q_1 &=(-1)^{n_2+1}x_1,\quad q_2=(-1)^{n_1+n_3+1}x_2,\quad q_3 =\frac{q_0}{x_1 x_2 }\,. 
\end{align}
\subsection*{Quiver $Q_1$}
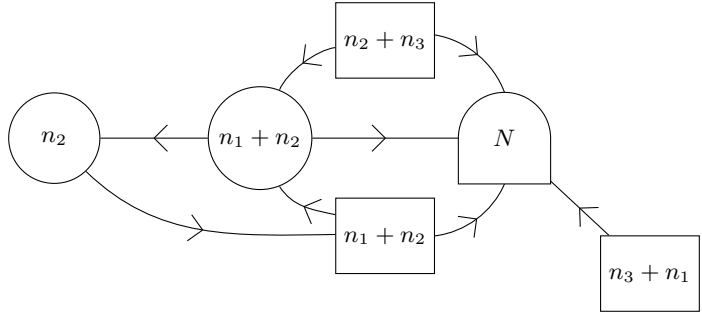
\begin{figure}[H]
\begin{center}
{\footnotesize
\begin{tikzpicture}[decoration={
markings,
mark=at position 0.5 with {\draw (-5pt,-5pt) -- (0pt,0pt);
                \draw (-5pt,5pt) -- (0pt,0pt);}}]
  \matrix[row sep=1mm,column sep=3mm] {
     &\node{$\phantom{n_3+}$}; & & \node[](BfNW)[squashedflavor] {$n_2+n_3$};  && \\      
     \node(Bg1)[gauge] {$n_2$};  & & \node(Bg2)[gauge] {$n_1+n_2$}; & &\node(Bgf)[gaugedflavor]{$N$};&\\
      & & &\node[](BfSW)[squashedflavor] {$n_1+n_2$};  &&\node[xshift=+10mm,yshift=-5mm](BfSE)[squashedflavor] {$n_3+n_1$}; \\
      };
\graph[edges={postaction={decorate}}]{
(Bg2)--(Bg1);(Bg2)--(Bgf);
(BfNW)--[bend right,looseness=.7](Bg2);(BfNW)--[bend left,right anchor=north,looseness=.9](Bgf);
(Bg1)--[bend right,in=+195](BfSW);(BfSW)--[bend left,out=+15](Bg2);(BfSW)--[bend right,left anchor=east,right anchor=south,looseness=.9](Bgf);
(BfSE)--(Bgf);
};
\end{tikzpicture}
}
\label{Q2app}
\end{center}
\end{figure}
The twisted superpotential is
\begin{align}
\label{WQ2y}
\mathcal{W}_{Q_1}(y)=&\log{y_1}\sum_{s\in {\mathcal N_2}}\sigma^{(1)}_s+\log{y_2}\sum_{s\in {\mathcal N}_1\cup {\mathcal N}_2}\sigma^{(2)}_s\cr
&-\sum_{t\in {\mathcal N}_1\cup {\mathcal N}_2}\sum_{s\in {\mathcal N_2}}\varpi(\sigma^{(2)}_t-\sigma^{(1)}_s)
-\sum_{s\in {\mathcal N}_1\cup {\mathcal N}_2}\langle\text{Tr}\,\varpi(\sigma^{(2)}_s-\Phi)\rangle\cr
&-\sum_{i\in {\mathcal F_1}}\sum_{s\in {\mathcal N_2}}\varpi(\sigma^{(1)}_s-m_i)
-\sum_{i\in \mathcal F_1\cup\mathcal F_2}\sum_{s\in {\mathcal N}_1\cup {\mathcal N}_2}\varpi(m_i-\sigma^{(2)}_s)\,.
\end{align}
The chiral ring equations are:
\begin{align}
\label{3nodeQ2TCR}
G_2 (\sigma_s^{(1)}) &= \frac{( -1 )^{n_1+n_2}}{ y_1} B_1 (\sigma_s^{(1)})\quad\text{for}\quad s\in {\mathcal N_2}\,,\cr
(1+q_0)\widehat P(\sigma_s^{(2)})&=( -1 )^{n_1+n_3}\Big( y_2 \frac{B_1(\sigma_s^{(2)}) B_2(\sigma_s^{(2)})}{G_1(\sigma_s^{(2)})} + \frac{q_0}{y_2} B_3(\sigma_s^{(2)})G_1(\sigma_s^{(2)}) \Big)\cr
&\hspace{7cm}\text{for}\,\, s\in {\mathcal N}_1\cup {\mathcal N}_2\,.
\end{align}
We solve the equations about the following vacuum:
\begin{align}
\sigma_s^{(1)} &= a_{s}  \quad\text{for}\quad s\in {\mathcal N_2}\,, \cr
\sigma_s^{(2)} &=  a_{s}  \quad\text{for}\quad s\in {\mathcal N}_1\cup {\mathcal N}_2\,.
\end{align}
Then at 1-instanton the twisted superpotential evaluated on the solution is:
\begin{align}
\mathcal{W}\big|_{\sigma_\star}&=\frac{(-1)^{n_1+n_2+1}}{y_1}\sum_{s\in\mathcal{N}_2}\frac{B_1(a_s)}{P_1(a_s)P'_2(a_s)}
+(-1)^{n_2+n_3+1}y_1\,y_2\sum_{s\in\mathcal{N}_2}\frac{B_2(a_s)}{P'_2(a_s)P_3(a_s)}\cr
&
\hspace{.5cm}+(-1)^{n_1+n_3+1}\frac{q_0}{y_2}\sum_{s\in\mathcal{N}_1}\frac{B_3(a_{s})}{P_3(a_s)P'_1(a_s)}\,.
\end{align}
This matches the 1-instanton results from localization using the contour $(-+-)$ and the map:
\begin{align}
q_1=\frac{(-1)^{n_2+1}}{y_1},\quad q_2=(-1)^{n_3}y_1\,y_2,\quad q_3 &=(-1)^{n_1+1}\frac{ q_0}{ \,y_2}\,.
 \end{align}

\subsection*{Quiver $Q_2$}

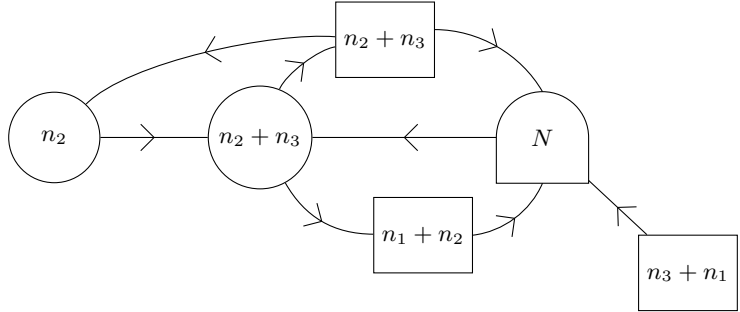
\begin{figure}[H]
\begin{center}
{\footnotesize
\begin{tikzpicture}[decoration={
markings,
mark=at position 0.5 with {\draw (-5pt,-5pt) -- (0pt,0pt);
                \draw (-5pt,5pt) -- (0pt,0pt);}}]
  \matrix[row sep=1mm,column sep=3mm] {
    &\node{$\phantom{n_3+}$}; & & \node[xshift=-5mm](CfNW)[squashedflavor] {$n_2+n_3$};  && \\      
     \node(Cg1)[gauge] {$n_2$};  & & \node(Cg2)[gauge] {$n_2+n_3$}; & &\node(Cgf)[gaugedflavor]{$N$};&\\
      & & &\node[](CfSW)[squashedflavor] {$n_1+n_2$};  &&\node[xshift=+10mm,yshift=-5mm](CfSE)[squashedflavor] {$n_3+n_1$}; \\
      };
\graph[edges={postaction={decorate}}]{
(Cg1)--(Cg2);(Cgf)--(Cg2);
(CfNW)--[bend right,out=-20,looseness=.6](Cg1);(Cg2)--[bend left,looseness=.6](CfNW);(CfNW)--[bend left,right anchor=north,looseness=.9](Cgf);
(Cg2)--[bend right](CfSW);(CfSW)--[bend right,left anchor=east,right anchor=south,looseness=.9](Cgf);
(CfSE)--(Cgf);
};
\end{tikzpicture}
}
\label{Q3app}
\end{center}
\end{figure}
The twisted superpotential is
\begin{align}
\label{WQ3z}
\mathcal{W}_{Q_2}(z)=&\log{z_1}\sum_{s\in {\mathcal N_2}}\sigma^{(1)}_s+\log{z_2}\sum_{s\in {\mathcal N}_2\cup {\mathcal N}_3}\sigma^{(2)}_s\cr
&-\sum_{t\in {\mathcal N}_2\cup {\mathcal N}_3}\sum_{s\in {\mathcal N_2}}\varpi(\sigma^{(1)}_s-\sigma^{(2)}_t)
-\sum_{s\in {\mathcal N}_2\cup {\mathcal N}_3}\langle\text{Tr}\,\varpi(\Phi-\sigma^{(2)}_s)\rangle \cr
&-\sum_{i\in {\mathcal F_2}}\sum_{s\in {\mathcal N_2}}\varpi(m_i-\sigma^{(1)}_s)
-\sum_{i\in \mathcal F_1\cup\mathcal F_2}\sum_{s\in {\mathcal N}_2\cup {\mathcal N}_3}\varpi(\sigma^{(2)}_s-m_i)\,.
\end{align}
The chiral ring equations are:
\begin{align}
\label{3nodeQ3TCR}
G_2 (\sigma_s^{(1)}) &=(-1)^{n_2+n_3}z_1 B_2 (\sigma_s^{(1)})\quad\text{for}\quad s\in {\mathcal N_2}\,,\cr
(1+q_0)\widehat P(\sigma_s^{(2)})&= (-1)^{n_1+n_3}\,\Bigg( \frac{B_1(\sigma_s^{(2)}) B_2(\sigma_s^{(2)})}{z_2 G_1(\sigma_s^{(2)})} + q_0 \,z_2 \,B_3(\sigma_s^{(2)})G_1(\sigma_s^{(2)})\Bigg) \, \,\text{for}\,\,s\in {\mathcal N}_2\cup {\mathcal N}_3\,. \cr
\end{align}
We solve the equations about the following vacuum:
\begin{align}
\sigma_s^{(1)} &= a_{s}  \quad\text{for}\quad s\in {\mathcal N_2}\,, \cr
\sigma_s^{(2)} &=  a_{s}  \quad\text{for}\quad s\in {\mathcal N}_2\cup {\mathcal N}_3\,.
\end{align}
Then at 1-instanton the twisted superpotential evaluated on the solution is:
\begin{align}
\mathcal{W}\big|_{\sigma_\star}&=\frac{(-1)^{n_1+n_2}}{z_1z_2}\sum_{s\in\mathcal{N}_2}\frac{B_1(a_{s})}{P_1(a_s)P'_2(a_s)}
+(-1)^{n_2+n_3}z_1\sum_{s\in\mathcal{N}_2}\frac{B_2(a_{s})}{P'_2(a_s)P_3(a_s)}\cr
&\hspace{.5cm}+(-1)^{n_3+1}q_0\,z_2\sum_{s\in\mathcal{N}_3}\frac{B_3(a_{s})}{P'_3(a_s)P_1(a_s)}\,.
\end{align}
This matches the 1-instanton results from localization following the contour $(-++)$ and the map
\begin{align}
q_1 &=\frac{(-1)^{n_2}}{z_1\,z_2},\quad q_2=(-1)^{n_3+1}z_1,\quad q_3 =(-1)^{n_1+1} q_0 \,z_2\,.
\end{align}
\subsection*{Quiver $Q_3$}
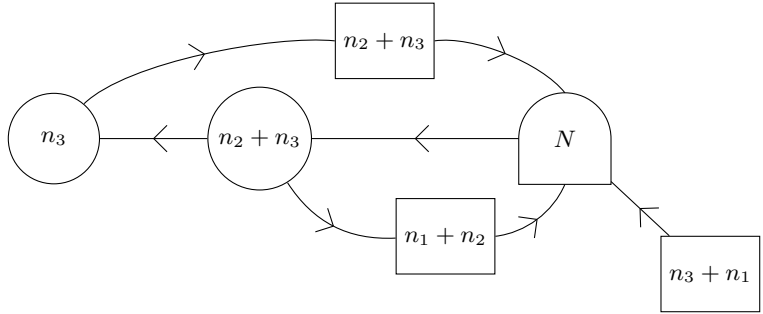
\begin{figure}[H]
\begin{center}
{\footnotesize
\begin{tikzpicture}[decoration={
markings,
mark=at position 0.5 with {\draw (-5pt,-5pt) -- (0pt,0pt);
                \draw (-5pt,5pt) -- (0pt,0pt);}}]
  \matrix[row sep=1mm,column sep=3mm] {
     &\node{$\phantom{n_3+}$}; & & \node[xshift=-8mm](DfNW)[squashedflavor] {$n_2+n_3$};  && \\      
     \node(Dg1)[gauge] {$n_3$};  & & \node(Dg2)[gauge] {$n_2+n_3$}; & &\node(Dgf)[gaugedflavor]{$N$};&\\
      & & &\node[](DfSW)[squashedflavor] {$n_1+n_2$};  &&\node[xshift=+10mm,yshift=-5mm](DfSE)[squashedflavor] {$n_3+n_1$}; \\
   };
\graph[edges={postaction={decorate}}]{
(Dgf)--(Dg2)--(Dg1);
(Dg1)--[bend left,looseness=.6,right anchor=west,in=-200](DfNW);(DfNW)--[bend left,left anchor=east,right anchor=north,looseness=.7](Dgf);
(Dg2)--[bend right](DfSW);(DfSW)--[bend right,left anchor=east,right anchor=south,looseness=.9](Dgf);
(DfSE)--(Dgf);
};
\end{tikzpicture}
}
\label{Q4app}
\end{center}
\end{figure}
The twisted superpotential is
\begin{align}
\label{WQ4w}
\mathcal{W}_{Q_3}(w)=&\log{w_1}\sum_{s\in {\mathcal N_3}}\sigma^{(1)}_s+\log{w_2}\sum_{s\in {\mathcal N}_2\cup {\mathcal N}_3}\sigma^{(2)}_s\cr
&-\sum_{t\in {\mathcal N}_2\cup {\mathcal N}_3}\sum_{s\in {\mathcal N_3}}\varpi(\sigma^{(2)}_t-\sigma^{(1)}_s)
-\sum_{s\in {\mathcal N}_2\cup {\mathcal N}_3}\langle\text{Tr}\,\varpi(\Phi-\sigma^{(2)}_s)\rangle\cr
&-\sum_{i\in {\mathcal F_2}}\sum_{s\in {\mathcal N_3}}\varpi(\sigma^{(1)}_s-m_i)
-\sum_{i\in \mathcal F_1}\sum_{s\in {\mathcal N}_2\cup {\mathcal N}_3}\varpi(\sigma^{(2)}_s-m_i)\,.
\end{align}
The chiral ring equations are:
\begin{align}
\label{3nodeQ4TCR}
G_2 (\sigma_s^{(1)}) &=(-1)^{n_2+n_3}\frac{B_2 (\sigma_s^{(1)})}{w_1}\quad\text{for}\quad s\in {\mathcal N_3}\,,\cr
(1+q_0)\widehat P(\sigma_s^{(2)}) &= (-1)^{N}\,\Bigg( \frac{ G_1(\sigma^{(2)}_s)B_1(\sigma^{(2)}_s)}{w_2}+ q_0 \,w_2 \frac{\,B_2(\sigma_s^{(2)})\,B_3(\sigma_s^{(2)})}{G_1(\sigma_s^{(2)})}\Bigg) \,
\,\text{for}\,\, s\in {\mathcal N}_2\cup {\mathcal N}_3\,. \cr
\end{align}
We solve the equations about the following vacuum:
\begin{align}
\sigma_s^{(1)} &= a_{s}  \quad\text{for}\quad s\in {\mathcal N_3}\,, \cr
\sigma_s^{(2)} &=  a_{s}  \quad\text{for}\quad s\in {\mathcal N}_2\cup {\mathcal N}_3\,.
\end{align}
Then at 1-instanton the twisted superpotential evaluated on the solution is:
\begin{align}
\mathcal{W}\big|_{\sigma_\star}&=\frac{(-1)^{N+1}}{w_2}\sum_{s\in\mathcal{N}_2}\frac{B_1(a_s)}{P_1(a_s)P'_2(a_s)}
+\frac{(-1)^{n_2+n_3+1}}{w_1}\sum_{s\in\mathcal{N}_3}\frac{B_2(a_{s})}{P_2(a_s)P'_3(a_s)}\cr
&\hspace{.5cm}+(-1)^{n_1+1}q_0\,w_1\,w_2\sum_{s\in\mathcal{N}_3}\frac{B_3(a_s)}{P'_3(a_s)P_1(a_s)}\,.
\end{align}
This matches the 1-instanton results from localization following the contour $(--+)$ and the map
\begin{align}
q_1=\frac{(-1)^{n_2+n_3+1}}{\,w_2},\quad q_2=\frac{(-1)^{n_3+1}}{w_1},\quad q_3 =(-1)^{n_1+n_3} q_0\,w_1 \,w_2\,.
 \end{align}

\end{appendix}

\providecommand{\href}[2]{#2}\begingroup\raggedright\endgroup

\end{document}